\documentstyle[aps,epsfig,multicol]{revtex}
\def\breakon{\end{multicols}\widetext\vspace{-.2cm}
\noindent\rule{.48\linewidth}{.3mm}\rule{.3mm}{.3cm}\vspace{.0cm}}

\renewcommand{\vec}[1]{{\mathbf #1}}

\def\breakoff{\vspace{-.2cm}
\noindent
\rule{.52\linewidth}{.0mm}\rule[-.27cm]{.3mm}{.3cm}\rule{.48\linewidth}{.3mm}
\vspace{-.3cm}
\begin{multicols}{2}
\narrowtext}

\newcommand{\be}{\begin{equation}}
\newcommand{\ee}{\end{equation}}
\newcommand{\bea}{\begin{eqnarray}}
\newcommand{\eea}{\end{eqnarray}}
\newcommand{\HH}{{\cal H}}

\newcommand{\p}{\partial}
\newcommand{\s}{\sigma}
\newcommand{\la}{\langle}
\newcommand{\ra}{\rangle}

\newcommand{\lp}{\left(}
\newcommand{\rp}{\right)}
\renewcommand{\vec}[1]{{\bf #1}}

\begin{document}
\title{
Dynamical Selection in Emergent Fermionic 
Pairing}
\author{R.\,A. Barankov and L.\,S. Levitov}
\address{Department of Physics, Massachusetts Institute of Technology,
77 Massachusetts Ave, Cambridge, MA 02139}
\date{\today}
\maketitle
\begin{abstract}
We consider evolution 
of a Fermi gas in the presence of a time-dependent BCS interaction.
The pairing amplitude
in the emergent BCS state is found to be an oscillatory
function of time with predictable characteristics.
The interplay of linear instability of the unpaired
state and nonlinear interactions selects 
periodic soliton trains of a specific form,
described by the Jacobi elliptic function dn.
While the parameters of the soliton train,
such as the period, amplitude, and time lag,
fluctuate among different realizations, 
the elliptic function form remains robust.
The parameter variation is accounted for by
the fluctuations of particle distribution in the initial unpaired state.
\end{abstract}

\begin{multicols}{2}

\narrowtext
Nonequilibrium effects in superconducting systems are usually described using
the notion of local equilibrium 
and quasiparticle distribution\cite{Langenberg86,Tinkham}, 
embodied in the time-dependent
Ginzburg-Landau equation\cite{Gorkov68} 
or the kinetic equation\cite{Aronov_review}.
The recently studied problem 
of fast time dynamics in emergent fermionic pairing\cite{Barankov03}
describing the onset of BCS pairing triggered by 
an abrupt change of interaction
lies somewhat outside this classification. 
The new aspect of this problem is the absence of a meaningful notion
of quasiparticle spectrum. Instead, individual Cooper pair states
evolve in a coherent, collisionless 
fashion\cite{Volkov74,Schmid66,Abrahams66,Galperin81} \emph{independently}
of the order parameter.
The ensuing interesting many-body evolution
reflects the system inability to transit 
adiabatically between the unpaired and paired states
due to their vanishing overlap. The oscillatory mode \cite{Barankov03} 
can be viewed as Bloch precession of
the pair states in an effective magnetic field
defined in terms of the pairing function.
 
Tunable time-dependent pairing interaction can be realized experimentally in
ultracold Fermi gases\cite{DeMarco991,Truscott01,Loftus02,Ohara02}.
Fermionic pairing and
superfluidity in these systems have been demonstrated recently
using magnetically tunable Feshbach 
resonances\cite{Regal04,Kinast04,Zwierlein04}.
In these systems, 
by varying the magnetic field detuning from resonance,
both the strength and the sign of inter-particle interaction 
can be changed on a time scale shorter than the intrinsic times
of fermions, set by $E_F$ or the elastic collision rate.
In Refs.\cite{Barankov04,Andreev04,Szymanska04} 
the picture of time-dependent pairing \cite{Barankov03} 
was extended to the problem of resonant atom-molecule coupling
near a Feshbach resonance. 

Another interesting class of tunable systems are the hybrid 
semiconductor-superconductor (S/Sm/S) and SNS 
structures\cite{Akazaki96,Volkov96,Schapers,Morpurgo98,Kutchinsky99} 
with proximity-induced Josephson effect. 
In the S/Sm/S systems\cite{Akazaki96,Volkov96}, 
the superfluid density and Josephson critical current 
can be tuned by electric field
applied on the gates to the semiconductor. 
Similarly, the SNS structures\cite{Schapers,Morpurgo98,Kutchinsky99} 
are tunable by current applied to the normal region.


Superconductor dynamics is controlled
by several kinetic time scales\cite{Tinkham}. 
The most important for us is
the collisionless regime\cite{Volkov74,Schmid66,Abrahams66,Galperin81} 
characterized by the time
\be\label{eq:tau_Delta}
\tau_\Delta=\hbar/\Delta
\ee
with $\Delta$ the equilibrium BCS gap. The time $\tau_\Delta$
manifests itself in the BCS instability growth 
rate\cite{Schmid66,Abrahams66,Galperin81} at $T$ away from
$T_c$, as well as in the frequency of pairing amplitude 
oscillations\cite{Volkov74}. 
Another time scale is 
set by the Landau Fermi liquid elastic collision time,
$\tau_{\rm el}^{-1}\propto 
\max[\epsilon^2,T^2]$
with $\epsilon$ the energy relative to $E_F$.
For typical energy $\epsilon\simeq\Delta$, in the weak coupling regime
$\Delta\ll E_F$, we estimate 
$\tau_{\rm el}\gg\tau_\Delta$.
For slowly varying interaction, 
the kinetic equation\cite{Aronov_review},
or the time-dependent Ginzburg-Landau equation\cite{Gorkov68}
can be employed.
In contrast, here we are interested in the situation when
the interaction is turned on abruptly on the scale $\tau_\Delta$, 
so that the hierarchy of times is
\be\label{eq:time_hierarchy}
\tau_0\ll \tau_\Delta\ll \tau_{\rm el}
,
\ee
where $\tau_0$ describes the interaction time dependence.
In this case, the time interval $0<t\lesssim \tau_{\rm el}$ 
is controlled by coherent pair dynamics, while energy relaxation
can be ignored. 
This regime can be described by the truncated BCS Hamiltonian,
which accounts for coherent Cooper pair transitions 
$(\vec p,-\vec p)\to (\vec p',-\vec p')$, 
ignoring other, more slow, processes of quasiparticle
scattering. In this article we 
study the resulting collisionless evolution, using a combination 
of analytical and numerical methods.

The time domain picture of BCS pairing buildup, 
proposed in Ref.\cite{Barankov03},
involves dynamics of individual pair states, selfconsistently coupled to
the pairing amplitude. The latter exhibits unharmonic
oscillations, undamped at $t<\tau_{\rm el}$. 
While several recent publications agree in general with
this scenario\cite{Amin04,Andreev04,Szymanska04,Warner05}, 
some of the issues remain unsettled.
Ref.\cite{Amin04} describes a simulation 
of the collisionless BCS dynamics exhibiting 
a damped oscillatory time dependence,
with the damping of unspecified origin.
These results, as well as those of Ref.\cite{Yuzbahyan_simul}, 
are clearly different
from the behavior found in Ref.\cite{Barankov03}.
The most probable origin of the discrepancy, in our view, 
is due to the numerical method\cite{Amin04}.

Below we provide extensive numerical evidence
which confirms the conclusions of Ref.\cite{Barankov03}.
We show that the mean field solutions,
the soliton trains described by elliptic dn function,
are singled out by the BCS evolution, and appear for generic 
initial conditions in a stable and robust way.
Estimates of noise made in Sec.\ref{sec:noise}
show that the mean field theory approach holds when
the level spacing in a relevant volume is small enough.
Our analysis indicates that the mean field results describe 
the dynamics of $|\Delta|$ in a wide temperature range 
not only in a finite, but also, locally, in an infinite system.

The physical reason for a specific solution to be selected by pairing dynamics
is due to the properties of the BCS instability. 
Linearization over the unpaired state of a Fermi gas in a 
finite system, discussed in Sec.\ref{sec:instability}, 
shows that, modulo phase degeneracy, 
there is just one unstable normal mode
corresponding to perturbation growth, while other modes do not grow. 
As a result, although the initial state is perturbed by thermal 
fluctuations in a completely random fashion, the BCS dynamics amplifies 
only one specific perturbation leading to an oscillatory time dependence with 
predictable characteristics.

\section{BCS instability of the unpaired Fermi gas}
\label{sec:instability}
%
The evolution of the Fermi gas with time-varying pairing
coupling can be described by the BCS Hamiltonian
\be\label{eq:Hbcs}
\HH = \sum_{\vec p,\,\s} 
\epsilon_{\vec p} a^+_{\vec p,\s}a_{\vec p,\s}
-\frac{\lambda(t)}2 
\sum_{\vec p,\vec q} a^+_{\vec p\,\uparrow}a^+_{-\vec p\,\downarrow}a_{-\vec q\,\downarrow}a_{\vec q\,\uparrow}
\,,
\ee
where $a_{\vec p,\s}$, $a^+_{\vec p,\s}$ are the canonical fermion operators,
and $\s=\uparrow,\downarrow$ is generalized spin. 
The time dependence of $\lambda$, 
as well as the resulting time dependence 
of the system state, is assumed to be fast on the 
scale of quasiparticle elastic collisions
and  energy relaxation, $\tau_{\rm el}$, allowing us to ignore the latter and 
consider the coherent dynamics defined by (\ref{eq:Hbcs}).
The simplest time dependence which we shall be most interested in, is 
described by 
the coupling turned on abruptly, from $\lambda(t<0)=0$ to
$\lambda(t>0)=\lambda$.

The interaction switching, while abrupt and nonadiabatic, 
must also be gentle enough not to overheat the Fermi system. 
The analysis of energy production due to two-particle scattering
in the presence of time-dependent coupling, described in Appendix, 
obtains an estimate for the effective temperature
\be\label{eq:Teff_FL}
T^2_{\rm eff}=\lambda^2\nu^3/\tau_0^{3}
.
\ee
The `no overheat' condition
$T_{\rm eff}\ll\Delta_0$
can thus be stated as
$E_F\tau_0\gg (\lambda n/\Delta_0)^{2/3}$, which is compatible with the
nonadiabaticity requirement $\tau_0\ll \tau_{\Delta}$.

Our treatment of the problem (\ref{eq:Hbcs}) will focus on the time-dependent
generalization of the BCS state\cite{Barankov03}
\be\label{eq:Psi_bcs}
|\Psi(t)\ra = \prod_{\vec p}\lp u_{\vec p}(t)+v_{\vec p}(t)a^+_{\vec p,\uparrow}a^+_{-\vec p,\downarrow}\rp |0\ra
.
\ee
The Bogoliubov mean field approach,
which gives a state of the form (\ref{eq:Psi_bcs}),
relies on the `infinite range' form of
the pairing interaction in (\ref{eq:Hbcs})
owing to equal coupling strength of all  
$(\vec p,-\vec p)$, $(\vec q,-\vec q)$. 
Since the latter does not depend
on the system being in equilibrium,
one can introduce a time-dependent mean field
pairing function
\be\label{eq:Delta_BCS}
\Delta(t)=\lambda\sum_{\vec p}u_{\vec p}(t)v_{\vec p}^\ast(t)
.
\ee
The amplitudes $u_{\vec p}(t)$, $v_{\vec p}(t)$
can be obtained from the Bogoliubov-deGennes equation
\be\label{eq:Bogoliubov_deGennes}
i\p_t \lp\matrix{ u_{\vec p} \cr v_{\vec p}}\rp =
\lp\matrix{ \epsilon_{\vec p} & \Delta \cr \Delta^\ast & -\epsilon_{\vec p}}\rp
\lp\matrix{ u_{\vec p} \cr v_{\vec p}}\rp
,
\ee
solved together
with the selfconsistency
condition (\ref{eq:Delta_BCS}).

We recall that the unpaired state is a selfconsistent, 
albeit unstable, solution of 
Eqs.\,(\ref{eq:Bogoliubov_deGennes}),(\ref{eq:Delta_BCS})
with $\Delta=0$, $T=0$:
\be\label{eq:u0v0}
u^{(0)}_{\vec p}(t) = 
e^{-i\epsilon_{\vec p}t}\theta(\epsilon_{\vec p})
,\quad
v^{(0)}_{\vec p}(t) = e^{i\phi_\vec p}e^{i\epsilon_{\vec p}t}\theta(-\epsilon_{\vec p})
,
\ee
with $\phi_\vec p$ a random phase. 
The stability analysis~\cite{Abrahams66}
shows that
the deviation from the unpaired state grows as $\Delta(t)
\propto e^{\gamma t}e^{-i\omega t}$, with linearized amplitudes 
\be\label{eq:exp(gamma_t)}
\delta u_{\vec p}(t)=
\frac{\Delta(t) v^{(0)}_{\vec p}(t)}{i\gamma - 2\epsilon_{\vec p}+\omega}
,\quad
\delta v_{\vec p}(t)=
\frac{\Delta^\ast(t) u^{(0)}_{\vec p}(t)}{i\gamma + 2\epsilon_{\vec p}-\omega}
.
\ee
The growth exponent $\gamma$ and the constant $\omega$,
combined into a complex number $\zeta=\omega+i\gamma$, 
are defined by the selfconsistency condition of the linearized problem:
\be\label{eq:instability_exponent}
1=\lambda\sum_{\vec p}\frac{{\rm sgn}\,\epsilon_{\vec p}}{2\epsilon_{\vec p}-\zeta}
.
\ee
This equation has a pair of complex conjugate solutions 
$\zeta$, $\zeta^\ast$.
From the similarity between Eq.\,(\ref{eq:instability_exponent})
and the BCS gap equation one expects the exponent $\gamma$ to be
close to the BCS gap value
$\Delta_0$ in equilibrium,
in which case the time constant $\tau_{\Delta}=\gamma^{-1}$
for initial pairing buildup
is of the order of $\Delta_0^{-1}$. 

It will be convenient for us to introduce here
the constant density of states approximation,
\be\label{nu_W}
\nu(\epsilon_\vec p)=\cases{\nu_0, & $|\epsilon_\vec p|<\frac12 W$\cr
0, & else}
,
\ee
used throughout this article. 
In our numerical study we use a total $N\gg 1$ equally spaced
discrete states 
distributed evenly in a finite size band,
%
\be\label{eq:bandwidth}
-W/2\le \epsilon_{\vec p} \le W/2
,
\ee
with the level spacing $\delta\epsilon=W/N$.
Although somewhat artificial, 
the model (\ref{eq:bandwidth}) introduces
a convenient simplification due to the particle-hole symmetry.
In this case, the chemical potential is locked at $\epsilon=0$
independent of the interaction strength. As a result, the instability
problem (\ref{eq:instability_exponent_T})
possesses an eigenvalue with a purely imaginary 
$\zeta=i\gamma$, satisfying
\be\label{eq:instability_exponent_symm}
1=\lambda\sum_{\vec p}\frac{2|\epsilon_{\vec p}|}{4\epsilon_{\vec p}^2
+\gamma^2}
,
\ee
and $\omega=0$.
Interestingly, if the bandwidth $W$
is much larger
than the BCS gap $\Delta_0$ at $T=0$, 
the value of the exponent 
$\gamma$ obtained from (\ref{eq:instability_exponent}) 
coincides with $\Delta_0$.
Indeed, Eq.(\ref{eq:instability_exponent_symm}) in this case gives
\be\label{eq:N0_exponent}
1=2\lambda \nu_0\int_0^{W/2}\frac{2\epsilon d\epsilon}{4\epsilon^2+\gamma^2}
=\frac12 \lambda \nu_0\ln\frac{W^2+\gamma^2}{\gamma^2}
\ee
This equation is reminiscent of the BCS gap equation
\[
1=\frac12\lambda\int_{-W/2}^{W/2}\frac{d\epsilon}{\sqrt{\epsilon^2+\Delta_0^2}}
=\lambda\nu_0\sinh^{-1}\frac{W}{2\Delta_0}
,
\]
becoming identical to it 
in the weak coupling limit of fermion energy band wide compared to
the gap, $W\gg \gamma,\,\Delta_0$
[\emph{cf.} Eq.(\ref{eq:Delta+(W)}) below].
The condition $\omega=0$ is violated in the absence
of particle-hole symmetry.
The equality $\gamma=\Delta_0$ holds only at $T=0$.

Let us briefly discuss how the analysis is modified 
in the case of a discrete spectrum  (\ref{eq:bandwidth}). 
With the sum in Eq.(\ref{eq:instability_exponent})
running over $N$ levels,
we obtain a polynomial of order $N$, having 
total $N$ roots. Simple analysis shows that $N-2$ roots are real 
and the remaining two are a complex conjugate pair $\zeta$, $\zeta^\ast$
with values close to that obtained for continuous spectrum. 
Accordingly, 
only the normal modes corresponding to $\zeta$, $\zeta^\ast$ are relevant 
for the instability, while the other $N-2$ modes correspond to 
the perturbations which do not grow
and remain small.
This conclusion is valid only at the times described by the 
linearized BCS dynamics. The situation at longer times,
which is more delicate, will be discussed below.

\begin{figure}[t]
\centerline{
\begin{minipage}[t]{7in}
\centering
\includegraphics[width=3.5in]{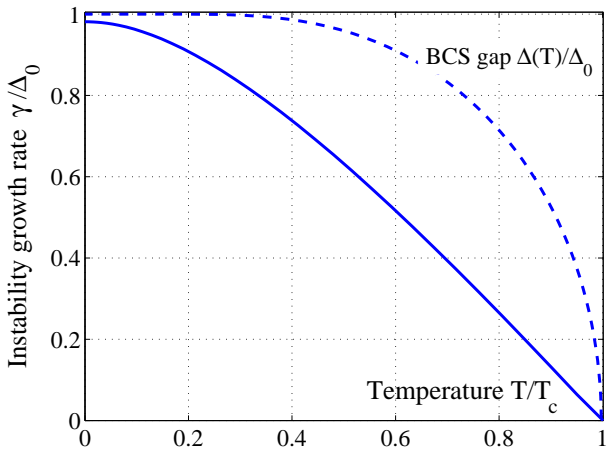}
\end{minipage}
}
\vspace{0cm}
\caption[]{
Temperature dependence of the BCS instability growth rate $\gamma$
as obtained from Eq.(\ref{eq:instability_exponent_T}) 
for the constant density of states model
(\ref{nu_W}), (\ref{eq:bandwidth}) with the coupling 
$\lambda$ such that $\Delta_0/W=1/5$.
Note that $\gamma$ coincides with the BCS gap $\Delta_0$ at $T=0$, 
up to a correction small as $(\gamma/W)^2$ at large bandwidth $W$. 
Near $T_c=\pi e^{-C}\Delta_0\simeq 1.764\Delta_0$,
the exponent $\gamma$ vanishes linearly in $T_c-T$.
}
\label{fig:gamma_temp}
\end{figure}

Let us now consider the instability at finite temperature.
The initial state describing
Fermi gas at $T>0$ can be tentatively chosen as
\be\label{eq:Psi_bcs_T}
\prod_{\vec p}\lp u_{\vec p}+c_\vec p a^+_{\vec p,\uparrow}
+c'_\vec p a^+_{-\vec p,\downarrow}
+v_{\vec p}a^+_{\vec p,\uparrow}a^+_{-\vec p,\downarrow}\rp |0\ra
,
\ee
where $u_\vec p$, $c_\vec p$, $c'_\vec p$, $v_{\vec p}$
equal zero or one depending on the occupancy:
$(u_{\vec p},c_{\vec p},c'_{\vec p},v_{\vec p})=(1,0,0,0),...,(0,0,0,1)$. 
The average values are given by occupation probabilities:
\[
\overline{|u_{\vec p}|^2}\!=\! (1-n_{\vec p})^2
,\ \ \overline{|v_{\vec p}|^2}\!=\! n_{\vec p}^2,\ \
\overline{|c_\vec p|^2}\!=\!\overline{|c'_\vec p|^2}\!=\!n_\vec p(1-n_\vec p), 
\]
where
$n_{\vec p}=1/(e^{\beta\epsilon_{\vec p}}+1)$ is the Fermi function, and
the quantities $n_{\vec p}^2$, $n_{\vec p}(1-n_{\vec p})$ and 
$(1-n_{\vec p})^2$
describe double, single and zero 
occupancy probability of the two-fermion 
states $(\vec p, -\vec p)$ in the unpaired system.
As we argue below, while the product state (\ref{eq:Psi_bcs_T}), 
written as a finite temperature generalization of (\ref{eq:Psi_bcs}), 
is not the most general fermion state, it is adequate
for our problem.

The product state (\ref{eq:Psi_bcs_T})
is suitable for simulation, since
Bogoliubov-deGennes dynamics (\ref{eq:Bogoliubov_deGennes}) couples
only $u_\vec p$ and $v_\vec p$ independently for each $\vec p$,  
preserving the product form. 
At the same time, the parts of (\ref{eq:Psi_bcs_T}) with single occupancy 
are decoupled from the collisionless pair 
dynamics (\ref{eq:Hbcs}). 
Indeed, the Hamiltonian (\ref{eq:Hbcs}) gives zero when applied to 
singly occupied pair states $a^+_{\vec k,\uparrow}|0\ra$, 
$a^+_{-\vec k,\downarrow}|0\ra$, irrespective of the occupancy of other
pair states. 
One can thus identify a subspace in the full Hilbert space,
which is spanned by all combinations of pair 
states of occupancy zero and two. The latter have the form
\be\label{eq:state_general}
|\psi\ra_{\rm general}={\prod_{\vec k}}' a^+_{\vec k,\uparrow}a^+_{-\vec k,\downarrow}|0\ra,
\ee
with the product taken over the states $(\vec k,-\vec k)$ 
of occupancy two whereby the states of occupancy one are excluded from the 
vacuum $|0\ra$.
The states (\ref{eq:state_general}) are mapped 
by (\ref{eq:Hbcs}) onto the states of a similar form,
thereby defining a full representation of the Hamiltonian.

Fortunately, one can bypass the combinatorics of 
(\ref{eq:state_general}) and simplify the state by employing 
Bogoliubov mean field approach which is exact for the 
Hamiltonian (\ref{eq:Hbcs}). In the mean field framework,
the general state is replaced by a product
state. Indeed, since the pairing amplitude $\Delta(t)$, 
describing cumulative effect of all pairs, is a c-number,
the dynamics (\ref{eq:Bogoliubov_deGennes}) does not generate
correlations between different pair states. 
At the same time, any correlations present
at $t=0$ are dephased by the dynamics itself, described by
time-dependent $2\times 2$ evolution matrices, different for each pair 
$(\vec p,-\vec p)$.
This argument, which will be refined in  Sec.\ref{sec:spin_1/2},
allows to replace the general states (\ref{eq:state_general}) 
by a simpler state of the product form (\ref{eq:Psi_bcs_T})
with $c_\vec p=c'_\vec p=0$. 
On a mean field level, the correlations between different
$(\vec p,-\vec p)$, $(\vec k,-\vec k)$ do not matter.

The effect of Pauli blocking which eliminates
the states of occupancy one can be taken into account
by going back to the $T=0$ state (\ref{eq:Psi_bcs})
with $u_{\vec p}$ and $v_{\vec p}$ chosen as
\be\label{eq:u0v0_T}
u^{(0)}_{\vec p}(t) = e^{-i\epsilon_{\vec p}t}(1-n_{\vec p})
,\quad
v^{(0)}_{\vec p}(t) = e^{i\phi_\vec p}e^{i\epsilon_{\vec p}t}n_{\vec p}
,
\ee
with random phase $\phi_\vec p$. 
The reduced norm $|u_\vec p|^2+|v_\vec p|^2=n^2_\vec p+(1-n_\vec p)^2<1$
reflects that at finite temperature some pair states, populated
by just one particle, are decoupled from the dynamics.
Near the Fermi level, at $|\epsilon_\vec p|\ll T$, the two-particle states
with double or zero occupancy have the probability
$1/4$ each, so that 
$|u_\vec p|^2+|v_\vec p|^2=1/2$. 
Outside this interval, $|\epsilon_{\vec p}|\geq T$, the blocking 
is practically absent and the norm approaches one.
Below we shall use the state (\ref{eq:Psi_bcs})
with 
$u_\vec p$, $v_\vec p$ given by
(\ref{eq:u0v0_T}) as initial condition
for the simulation.

The choice (\ref{eq:u0v0_T}), while somewhat \emph{ad hoc}, 
has an additional advantage over
choosing $u_\vec p=1,0$, $v_\vec p=0,1$. 
With both $u_\vec p$ and $v_\vec p$ nonzero with random relative phase,
the state (\ref{eq:Psi_bcs})
exhibits pairing fluctuations, providing a ``seed'' for the BCS instability
in the simulation. 
The results of the latter indicate that this choice of the initial state 
is general enough for the instability to fully play out. 
As we shall see in Sec.\ref{sec:spin_1/2}, where a spin $1/2$ formulation
of BCS dynamics is discussed, $u_\vec p$ and $v_\vec p$
can be understood as a two-component spinor. This means that the state
of the entire system
can indeed be chosen in the product form, with the initial 
values $u_\vec p$, $v_\vec p$ having random modulus, not just phase. 
This difference, however, is inessential, since the modulus of 
$u_\vec p$, $v_\vec p$
is quickly randomized by the dynamics itself.

Returning to the analysis of the instability, 
linearization of (\ref{eq:Bogoliubov_deGennes})
over the finite temperature state
(\ref{eq:u0v0_T}) obtains a time-dependent perturbation 
of the form
Eq.(\ref{eq:exp(gamma_t)}), and a generalization of
Eq.(\ref{eq:instability_exponent}), with 
${\rm sgn}\,\epsilon_{\vec p}$ replaced by 
\[
\overline{|u_{\vec p}^{(0)}|^2}-\overline{|v_{\vec p}^{(0)}|^2}=1-2n_{\vec p}
=\tanh\frac12\beta\epsilon_{\vec p}.
\]
The resulting equation,
\be\label{eq:instability_exponent_T}
1=\lambda\sum_{\vec p}\frac{1-2n_{\vec p}}{2\epsilon_{\vec p}-\zeta}
,\quad
\Delta(t) \propto e^{-i\zeta t}
,
\ee
has nonzero solutions 
$\zeta$, $\zeta^\ast$
below the BCS transition, $T<T_c$, which are purely imaginary,
$\zeta,\,\zeta^\ast=\pm i\gamma$, in the case of particle-hole symmetry
(\ref{nu_W}), (\ref{eq:bandwidth}).
The growth exponent $\gamma$
vanishes as $T\to T_c$, as shown in Fig.\ref{fig:gamma_temp}. 
This can be verified by 
noting that Eq.(\ref{eq:instability_exponent_T})
yields infinitesimally small $\gamma$ at $T=T_c$,
since at $\zeta=0$ it becomes identical 
to the BCS equation for $T_c$.

Having established the form of the unstable mode
(\ref{eq:exp(gamma_t)}) obtained from linearization, let us now
estimate the time range for which this analysis is accurate. 
The denominator in Eq.(\ref{eq:exp(gamma_t)}) is larger than the numerator
as long as $\Delta(t)\lesssim \gamma$, 
with $\gamma\approx \Delta(T)$, the BCS gap.
The initial value $\eta=\Delta(t=0)$, nonzero due to fluctuations in the 
unpaired Fermi system, is small in $1/N$:
\be\label{eq:Delta_fluct}
\eta=\lambda\sum_{\vec p}u_\vec p v_{\vec p}^\ast
=\lambda\sum_{\vec p}n_\vec p (1-n_\vec p)e^{i\phi_\vec p}
.
\ee
The sum (\ref{eq:Delta_fluct}) is controlled by about $T/\delta\epsilon$ terms 
with $|\epsilon_\vec p|\lesssim T$, uncorrelated in phase.
From the central limit theorem argument, 
we estimate $\eta\simeq \lambda\nu\sqrt{T\delta\epsilon}$ 
by order of magnitude.
The condition $\Delta(t)\lesssim \gamma$,
with exponentially growing $\Delta(t)=\eta e^{\gamma t}$,
defines the time interval $0<t\lesssim \gamma^{-1}\ln(\gamma/\eta)$
in which the evolution is 
described by the linearized problem.

Our aim will be
to gain insight in the behavior at later times.
We rely on the nondissipative character of the Bogoliubov-deGennes
dynamics (\ref{eq:Bogoliubov_deGennes}),(\ref{eq:Delta_BCS}),
manifest, for instance, in the energy
$E=\la \Psi(t)|\HH |\Psi(t)\ra$ conservation throughout the evolution.
Since for the unpaired state the energy exceeds its value 
in the ground state
by the BCS condensation energy, the equilibrium cannot be reached
without collisions.
Eqs.\,(\ref{eq:Bogoliubov_deGennes}),(\ref{eq:Delta_BCS}) hold
at times shorter than quasiparticle
thermalization time $\tau_{\rm el}$.
For temperatures away from $T_c$,
the time $\tau_{\rm el}$, evaluated at $\epsilon\simeq \Delta_0$, 
is long compared to the instability growth time,
\be\label{eq:Tau_el>>gamma}
\tau_{\rm el}\gg \gamma^{-1}
.
\ee
This means that the linear instability phase
is followed by a long {\it collisionless
nonlinear phase}.
Below we show that the evolution governed by
Eqs.\,(\ref{eq:Bogoliubov_deGennes}),(\ref{eq:Delta_BCS})
is described by soliton-like pulses in $\Delta(t)$. 
We obtain a family of exact solutions
of the form of single solitons and soliton trains, and compare it with simulations.

We briefly note that the importance of coherent dynamics of individual
Cooper pairs (\ref{eq:Psi_bcs})
in the time evolution of a paired state has been understood 
a long time ago
in the context of the discussion of the validity of the time-dependent
Ginzburg-Landau equation approach~\cite{Schmid66,Abrahams66,Gorkov68}. 
It has been pointed out
by Gorkov and Eliashberg~\cite{Gorkov68} that
the time-dependent
pairing function $\Delta$ is generally insufficient to describe
the evolution.
For such a description to be consistent,
the pair breaking and energy relaxation 
must be fast compared to the time scale $\tau_\Delta$ of change
of $\Delta$. This can be realized only 
close enough to the transition, or in superconductors with magnetic 
impurities~\cite{Gorkov68}, 
where the inequality (\ref{eq:Tau_el>>gamma}) is violated. 
Except these special situations,
however, the Cooper pairs
{\it are not slaved} to the time-dependent $\Delta(t)$,
and the dynamics of each 
pair has to be treated individually, via
Eq.\,(\ref{eq:Bogoliubov_deGennes}).

\section{Bogoliubov-deGennes equation as a Bloch equation}
\label{sec:Bloch_eqn}

To analyze the nonlinear BCS dynamics 
we reformulate the Bogoliubov approach,
bringing it to a form amenable to analytic and numerical treatment.
We show that the evolution 
of time-dependent amplitudes
$u_{\vec p}(t)$, $v_{\vec p}(t)$, governed by the 
Bogoliubov-deGennes equation (\ref{eq:Bogoliubov_deGennes}) 
with the selfconsistency condition (\ref{eq:Delta_BCS}), 
can be cast in the form of a Bloch equation for auxiliary variables. 
This is achieved by by introducing a new set of variables,
\be\label{eq:gf}
f_\vec p=2u_{\vec p}v_{\vec p}^\ast
,\quad
g_\vec p=|u_{\vec p}|^2-|v_{\vec p}^\ast|^2
.
\ee
Applied to the quantities (\ref{eq:gf}), 
Eq.(\ref{eq:Bogoliubov_deGennes}) gives a system of coupled equations:
\be\label{eq:VolkovKogan_eqs}
\frac{df_\vec p}{dt}=-2i\epsilon_\vec p f_\vec p +2i\Delta g_\vec p
,\quad
\frac{dg_\vec p}{dt}=i\Delta^\ast f_{\vec p}-i \Delta f_{\vec p}^\ast
.
\ee
These are nothing but the Gorkov equations \cite{Gorkov_eqs}
for the Green's functions
$G$ and $F$. In the form (\ref{eq:VolkovKogan_eqs}),
using momentum-dependent quantities $f_\vec p$, $g_\vec p$,
these equations were first written by Volkov and Kogan\cite{Volkov74}.

The dynamics (\ref{eq:VolkovKogan_eqs}) is to be supplemented by 
the Gorkov selfconsistency relation \cite{Gorkov_eqs},
\be\label{eq:Gorkov}
\Delta=\frac{\lambda}2\sum_\vec p f_\vec p
,
\ee
which defines $\Delta(t)$ through the values of evolving $f_\vec p(t)$. 
The initial conditions
\be\label{eq:fg_initial}
f_\vec p^{(0)}=e^{i\phi_\vec p}(1-n_\vec p)n_\vec p
,\quad
g_\vec p^{(0)}=1-2n_\vec p
,
\ee
correspond to the unpaired Fermi gas state
(\ref{eq:u0v0_T}).
 
To gain insight in the behavior of Eqs.(\ref{eq:VolkovKogan_eqs}),
it will be convenient to introduce the Bloch representation
\be\label{eq:Bloch_vector}
\vec r_\vec p=
(r_1,r_2,r_3)_\vec p
,\quad
r_1+ir_2=f_\vec p
,\quad
r_3=g_\vec p
.
\ee
The norm of $\vec r_\vec p$ is given by
\be\label{eq:norm_Q}
|\vec r_\vec p|=
\sqrt{ |f_\vec p|^2+g_\vec p^2}=n^2_\vec p+(1-n_\vec p)^2
.
\ee
Remarkably, after rewriting Eq.(\ref{eq:VolkovKogan_eqs}) 
in terms of $\vec r_{\vec p}$, it assumes the form of a Bloch equation:
\be\label{eq:Bloch_r}
\frac{d\vec r_{\vec p}}{dt}=2\vec b_\vec p\times \vec r_{\vec p}
,
\ee
where the ``magnetic field''
$\vec b_\vec p=-(\Delta',\Delta'',\epsilon_\vec p)$
has time-dependent $x$ and $y$ component satisfying
the selfconsistency condition (\ref{eq:Gorkov}).
The norm of the Bloch vectors $\vec r_\vec p$, given by
Eq.(\ref{eq:norm_Q}), is less than one, 
which describes the effect of Pauli blocking, i.e.
decoupling of the states with single occupancy from the BCS dynamics
(see Sec.\ref{sec:instability}).

\begin{figure}[t]
\centerline{
\begin{minipage}[t]{3.5in}
\centering
\includegraphics[width=3.5in]{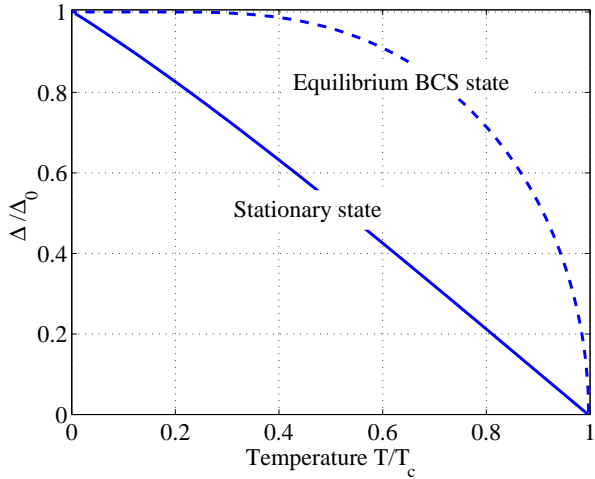}
\end{minipage}
}
\vspace{0cm}
\caption[]{
Temperature dependence of the pairing amplitude
$\Delta$
for the stationary state
(\ref{eq:stationary_z}), (\ref{eq:stationaryT})
obtained from the unpaired state by adiabatic increase
of coupling.
The equilibrium
BCS gap is shown for comparison.
}
\label{fig:delta_stat}
\end{figure}

Before exploring the dynamics, let us inspect the stationary states
of Eq.(\ref{eq:Bloch_r}). The latter are described by the
Bloch spins aligned with the 
magnetic field axis,
$\vec l_\vec p=-\vec b_\vec p/|\vec b_\vec p|$. In this case we have 
\be\label{eq:stationary_z}
(l_1+il_2)_\vec p=\frac{\Delta}{\sqrt{\epsilon^2_\vec p+|\Delta|^2}}
,\quad
l_{3,\vec p}=\frac{\epsilon_\vec p}{\sqrt{\epsilon^2_\vec p+|\Delta|^2}}
,
\ee
and the selfconsistency condition (\ref{eq:Gorkov}),
which determines the stationary value of $\Delta$,
takes the form
\be\label{eq:stationaryT}
1=\frac{\lambda}{2}\sum_{\vec p}\frac{\tanh\frac12\beta|\epsilon_\vec p|}{\sqrt{\epsilon^2_\vec p+|\Delta|^2}}
.
\ee
The numerator, $\tanh(\frac12\beta|\epsilon_\vec p|)=|1-2n_\vec p|$,
is the length of the Bloch vector $\bar{\vec r}_\vec p$
\emph{averaged}
over a group of levels with nearly equal energies, 
$|\epsilon_\alpha-\epsilon_\beta|\ll \Delta$.
The averaging, applied to the initial $\vec r_\vec p$
values (\ref{eq:fg_initial}), 
eliminates the transverse part
of (\ref{eq:fg_initial}), containing random phases $\theta_\vec p$,
while leaving the longitudinal part intact. 
The Bloch dynamics is unitary with respect 
to each Bloch vector $\vec r_\vec p$ and, in particular, 
is linear and preserves the norm. As a result,
the averaged vectors $\bar{\vec r}_\vec p$ will evolve according to
the same Bloch equations, albeit having a smaller norm
$|\bar{\vec r}_\vec p|=|1-2n_\vec p|< |\vec r_\vec p|$.

The equation (\ref{eq:stationaryT}) is different from the BCS gap equation
which contains $\tanh(\frac12\beta(\epsilon_\vec p^2+\Delta^2)^{1/2})$
instead of $\tanh(\frac12\beta|\epsilon_\vec p|)$,
except $T=0$, when
these equations coincide since $\tanh=\pm1$ in both cases.
Thus at temperatures $0<T<T_c$, as Fig.\ref{fig:delta_stat} illustrates,
Eq.(\ref{eq:stationaryT}) predicts the stationary value of 
$\Delta$ below the BCS gap scale. The temperature at which
$\Delta$ vanishes coincides with the BCS critical temperature,
since the condition (\ref{eq:stationaryT}) at $\Delta=0$
is identical to the BCS equation for $T_c$.

To clarify the character of the 
states (\ref{eq:stationary_z}), (\ref{eq:stationaryT}), 
one can make the following observations.
The only difference here from the BCS theory 
is due to incomplete 
equilibrium, owing to the singly occupied states being
Pauli-blocked from the dynamics controlled by (\ref{eq:Hbcs}).
Indeed, the truncated BCS Hamiltonian (\ref{eq:Hbcs}) accounts only for the 
collisionless pair dynamics,
but not for single particle 
scattering and relaxation. 
The latter processes have characteristic 
rates set by the two-particle collisions,
$1/\tau_{\rm el}$.
Since $\tau_{\rm el}$ is larger than 
$\tau_\Delta=\gamma^{-1}$,
the approach 
accounting only for the
coherent dynamics, but not for relaxation, is valid in 
a relatively large time interval 
$0<t\lesssim\tau_{\rm el}$. 

The stationary nonequilibrium states
(\ref{eq:stationary_z}), (\ref{eq:stationaryT}) can be realized
when the coupling constant $\lambda$ increases as a function of time slowly 
on the scale of $\tau_\Delta$, i.e. the condition 
(\ref{eq:time_hierarchy}) is replaced by
\[
\tau_\Delta\ll \tau_0\ll \tau_{\rm el}
.
\]
In this case, each 
Bloch vector $\vec r_\vec p$ 
follows the direction of the field $\vec b_\vec p(t)$,
maintaining constant projection on the $\vec b_\vec p(t)$ axis
equal to $1-2n_\vec p$ on average. During the evolution,
the value $\Delta(t)$ is determined by the selfconsistency 
condition (\ref{eq:Gorkov}).
The amplitudes of the pair states with occupancy zero and two
thereby become slaved to the adiabatic dynamics $\Delta(t)$, evolving
according to 
(\ref{eq:stationary_z}), (\ref{eq:stationaryT}).
At the same time, the amplitudes with occupancy one
remain decoupled, and do not evolve at times $t<\tau_{\rm el}$.
Such a behavior can be seen as a result of the evolution 
which is simultaneously
adiabatic in the pair sector, and totally nonadiabatic 
in the single particle sector.

\section{Oscillatory solutions, analytical and numerical}
\label{sec:oscillations}

Here we consider the nonlinear BCS dynamics described by the Bloch equation
(\ref{eq:Bloch_r}) and the selfconsistency relation (\ref{eq:Gorkov})
at the times after the instability sets in.
Eq.\,(\ref{eq:Bloch_r})
is quite easy to simulate, since it is
linear in $\vec r_\vec p$ and is written for classical, 
rather than quantum variables. 
The initial state,
Eq.(\ref{eq:fg_initial}), describing free fermions at 
a finite temperature, is 
\be\label{eq:r_initial}
r_{1,\vec p}\!+\!i r_{2,\vec p}=
\frac{e^{i\phi_{\vec p}}}{\cosh(\textstyle{\frac12}\beta\epsilon_{\vec p})}
,\
r_{3,\vec p}=\tanh(\textstyle{\frac12}\beta\epsilon_{\vec p})
,
\ee
with uncorrelated phases, uniformly distributed
in the interval $0<\phi_{\vec p}<2\pi$.
This form of initial conditions corresponds to
the amplitudes $u_\vec p$, $v_\vec p$ of the form (\ref{eq:u0v0_T}).
To avoid confusion with temporal characteristics, 
such as the oscillation period
$T$, hereafter we shall use the inverse temperature $\beta$, 
unless explicitly stated otherwise.

The dynamics (\ref{eq:Bloch_r}), (\ref{eq:Gorkov})
was obtained from $3N$ coupled differential equations
with the initial conditions (\ref{eq:r_initial}),
with $N$ large enough to ensure proximity to the continual limit.
The numerics was executed using 
the Runge-Kutta method with precision $O(dt^5)$.
The time step $dt$ was varied over a range of values to test
numerical accuracy. 
We found that the step $dt=0.01/\Delta_0$ typically
provides sufficient precision over the time interval
of interest.

\begin{figure}[t]
\centerline{
\begin{minipage}[t]{3.5in}
\centering
\includegraphics[width=3.5in]{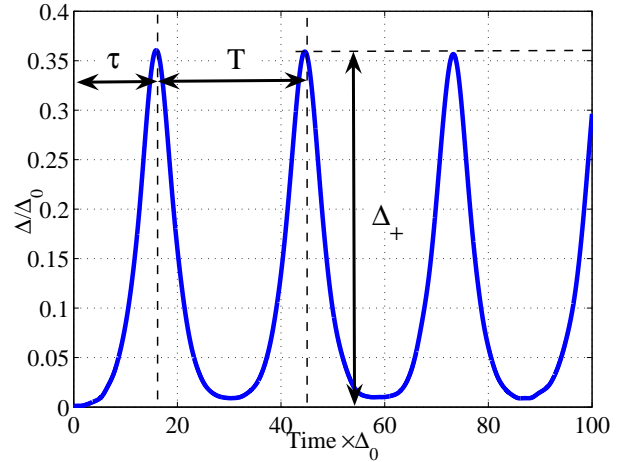}
\end{minipage}
}
\vspace{0cm}
\caption[]{
Time dependence of the pairing amplitude
$\Delta$ recorded from simulation
with $N=10^5$ states 
(\ref{nu_W}), (\ref{eq:bandwidth}) 
at temperature $T=0.7 T_c$
($\beta=2.5\Delta_0$) with the 
initial conditions (\ref{eq:r_initial}). 
The coupling constant
$\lambda$ was chosen to have the BCS gap
$\Delta_0=W/5$.
}
\label{fig:delta_sample}
\end{figure}

As Fig.\ref{fig:delta_sample} illustrates, 
a straightforward simulation generates a surprisingly regular, oscillatory
time dependence
$\Delta(t)$ which appears to be a periodic function of time.
After initial exponential growth, controlled by the instability discussed in
Sec.\ref{sec:instability}, we observe an essentially periodic time dependence,
characterized by equally spaced peaks of identical shape.
We can thus define the temporal period $T$, the time lag $\tau$,
and the amplitude which we denote by $\Delta_+$ 
for the reasons to become clear below. 
These notations are 
marked in Fig.\ref{fig:delta_sample}.

It might seem that
the particle-hole symmetry of the
density of states (\ref{nu_W}), (\ref{eq:bandwidth}) would enforce
zero chemical potential, regardless of BCS interaction strength.
In the simulation, however, we observe nonzero values of chemical potential
due to particle-hole imbalance caused by thermal fluctuations 
in the initial distribution $n_\vec p$. The complex pairing amplitude 
exhibits a phase growing linearly with time, 
$\Delta(t)=e^{-i\phi(t)}|\Delta(t)|$, $\phi(t)\propto\omega t$, 
with $\omega$ random in sign, constant for each realization,
equal twice the chemical potential. 
In addition, we observe noise superimposed on the linear time dependence
of the phase, which we shall discuss below at the end of this section.
As noted in Ref.\cite{Barankov03}, 
finite $\omega$ can be eliminated by a gauge transformation 
which shifts single particle energies by $\omega/2$, 
$\tilde\epsilon_\vec p=\epsilon_\vec p-\omega/2$, thereby making $\Delta$ real.
In the Bloch equation language, this is equivalent to considering the problem
in a Larmor frame rotating about the $\hat {\vec z}$ axis with frequency
$\omega$. Having this in mind, below we shall focus on the behavior of
the modulus $|\Delta|$.

Interestingly, the obtained time dependence $\Delta(t)$ can be fitted 
extremely accurately
to the analytic solution found in Ref.\cite{Barankov03}
for noiseless initial conditions. 
The latter is given by a Jacobi elliptic function,
periodic in time,
\be\label{eq:Delta=dn}
\Delta(t)=\Delta_+ {\rm dn\,}(\Delta_+(t-\tau),k)
,\quad
k^2=1-\Delta^2_-/\Delta^2_+
,
\ee
with $\Delta_+$ the amplitude, $\tau$ the time lag, and $\Delta_-$
the minimal value. 
We recall that the function $u={\rm dn\,}(x,k)$ 
is obtained by inversion of an elliptic integral:
\be\label{eq:elliptic_integral}
x=\int_u^1\frac{du'}{\sqrt{(1-u'^2)(k^2-1+u'^2)}}
\ee
The function (\ref{eq:Delta=dn}) satisfies the differential equation
\be\label{eq:dot_Delta_pm}
(d\Delta/dt)^2=(\Delta^2_+-\Delta^2)(\Delta^2-\Delta^2_-)
\ee
with $\Delta_\pm$ being the extremal values:
$\Delta_-\le\Delta(t)\le\Delta_+$. 
The period of the function (\ref{eq:Delta=dn})
is given by the complete elliptic integral of the 1st kind:
\be
T=\frac2{\Delta_+}K(k)=\frac2{\Delta_+}\int_0^{\pi/2}\frac{d\phi}{\sqrt{1-k^2\sin^2\phi}}
.
\ee
For sparse soliton trains, $T\Delta_+\gg1$, this expression simplifies
to $T=\frac2{\Delta_+}\ln(4\Delta_+/\Delta_-)$.

The time dependence of the Bloch vectors
$\vec r_\vec p(t)$ can be obtained from the ansatz
\be\label{eq:r_123_ansatz}
z_\vec p=A_\vec p\Delta(t)+iB_\vec p\dot\Delta(t)
,\quad
r_{3,\vec p}=C_\vec p\Delta^2-D_\vec p
.
\ee
Eqs.(\ref{eq:VolkovKogan_eqs}) are satisfied
by (\ref{eq:r_123_ansatz}) provided
$A_{\vec p}=2\epsilon_{\vec p}B_{\vec p}$ and $B_{\vec p}=-C_{\vec p}$.
The normalization condition $r_1^2+r_2^2+r_3^2=1$ thereby
turns into Eq.(\ref{eq:dot_Delta_pm}), the same for all $\vec p$,
yielding the relation 
between $C_{\vec p}$, $D_{\vec p}$ and $\Delta_\pm$:
\be\label{eq:CD_Delta}
\frac{D^2_{\vec p}-1}{C^2_{\vec p}}=\Delta^2_-\Delta^2_+
,
\quad
2\frac{D_{\vec p}}{C_{\vec p}}=4\epsilon_{\vec p}^2+\Delta^2_-+\Delta^2_+
.
\ee
Here $C_{\vec p}$ and $D_{\vec p}$, and likewise $A_{\vec p}$ and $B_{\vec p}$,
depend on $\epsilon_\vec p$, while the quantities
$\Delta_\pm$ are the same for all $\vec p$.

\begin{figure}[t]
\centerline{
\begin{minipage}[t]{3.5in}
\centering
\includegraphics[width=3.5in]{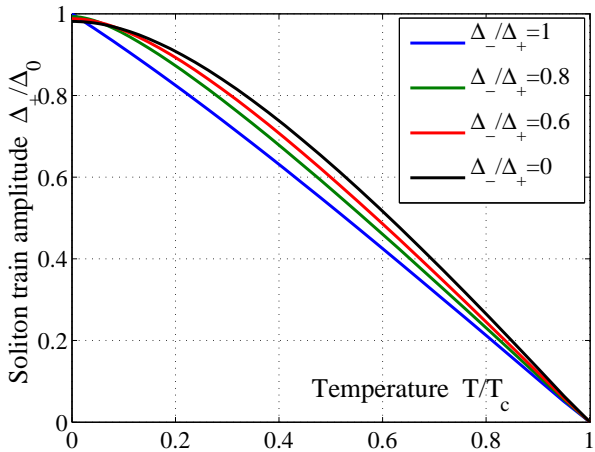}
\end{minipage}
}
\vspace{0cm}
\caption[]{
Temperature dependence of the soliton train amplitude
$\Delta_+$, obtained from the selfconsistency condition
(\ref{eq:Delta_01}) at different ratios $\Delta_-/\Delta_+$. Note that
at $\Delta_-=0$ the amplitude $\Delta_+$ equals the BCS instability growth
increment $\gamma$ (see Fig.\ref{fig:gamma_temp}),
while at $\Delta_-=\Delta_+$ the result for the stationary state
is reproduced (see Fig.\ref{fig:delta_stat}).
}
\label{fig:delta_temp}
\end{figure}

A special property of the ansatz (\ref{eq:r_123_ansatz}) which makes
it compatible with the selfconsistency condition (\ref{eq:Gorkov}),
is that $f_{\vec p}=A_\vec p\Delta(t)$ has the same time dependence
as the left hand side of Eq.(\ref{eq:Gorkov}). Therefore,
the selfconsistency will hold at all times provided that the quantities
$\Delta_\pm$ are chosen to satisfy
$1=\frac{\lambda}2\sum_\vec p |\bar{\vec r}_\vec p| A_\vec p$.
Here the averaging $\bar{\vec r}_\vec p$ over a group of levels 
with close energies is performed in the same way as in the derivation of
Eq.(\ref{eq:stationaryT}). (As above, the averaging is compatible
with unitary evolution due the linear character of Bloch dynamics.)
After substituting the 
expressions for $A_\vec p$ in terms of $\Delta_\pm$, 
and $\bar{\vec r}_\vec p=1-2n_\vec p$, we obtain
\be\label{eq:Delta_01}
1=\lambda\sum_{\vec p}
\frac{2\epsilon_\vec p \tanh(\frac12\beta\epsilon_\vec p)}{\lp
(4\epsilon_{\vec p}^2+\Delta^2_- +\Delta^2_+)^2-4\Delta^2_-\Delta^2_+\rp^{1/2}}
.
\ee
The role of this equation is similar to the BCS gap equation.
The only difference is that 
it fixes one of the two constants $\Delta_{\pm}$, leaving the other one free.

The motivation to consider this particular solution can be seen from 
the behavior of the elliptic integral (\ref{eq:elliptic_integral}) 
at $\Delta_-=0$. In this case, we obtain 
a single soliton 
\be\label{eq:cosh_soliton}
\Delta(t)=\frac{\Delta_+}{\cosh\Delta_+(t-t_0)}
.
\ee
At large negative time, Eq.(\ref{eq:cosh_soliton}) 
describes exponential growth of 
$\Delta$. Furthermore, Eq.(\ref{eq:Delta_01}) at $\Delta_-=0$
is identical to the condition (\ref{eq:instability_exponent_T})
for the instability growth rate, so that $\Delta_+=\gamma$.
Thus the single soliton solution (\ref{eq:cosh_soliton}) 
describes the nonlinear evolution 
following the linear instability regime. The nonmonotonic behavior
of $\Delta(t)$, first growing and then decreasing to zero, can be understood
as a result of energy mismatch of the BCS ground state and the 
unpaired state: Energy conservation in the collisionless dynamics
prevents system to evolve to the ground state with lower energy.

Remarkably, while these solutions appear to be very special, they are robust
in the presence of noise. Below we study the instability of 
Fermi gas at finite temperature, and find that the time
dependence survives thermal fluctuations
in the initial state. The reason for such a behavior is owing to the 
property of BCS instability, discussed in Sec.\ref{sec:instability}, 
to develop through a single unstable mode. As a result, only the 
fluctuation in the initial state along the unstable direction 
is amplified by BCS dynamics, while other fluctuations remain small,
providing a selection mechanism for the solutions (\ref{eq:Delta_01}).

Returning to the analysis of soliton trains (\ref{eq:Delta=dn}), we note that
the ratio $r=\Delta_-/\Delta_+$ controls the inter-soliton
time separation. Different regimes can be
qualitatively understood by noting that $\Delta$ varies in the
interval $\Delta_-\le \Delta(t) \le\Delta_+$.
For $r$ increasing from $0$ to $1$,
the soliton train period $T$ decreases, 
making the solitons overlap stronger and
gradually merge,
turning into weak harmonic oscillations with frequency $2\Delta_+$
as $\Delta_-$ approaches $\Delta_+$ 
(see Fig.\,1 of Ref.\cite{Barankov03}).

As a function of temperature,
the quantity $\Delta_+$ varies from the 
value close to the BCS gap $\Delta_0$ at $T=0$,
to zero at $T=T_c$ (see Fig.\ref{fig:delta_temp}).
At $\Delta_-\ll\Delta_+$, Eq.\,(\ref{eq:Delta_01}) turns into
Eq.\,(\ref{eq:instability_exponent})
which, as we found above, defines the amplitude of a single soliton
(\ref{eq:cosh_soliton}).
In the opposite limit,
$\Delta_-\to\Delta_+$, Eq.\,(\ref{eq:Delta_01})
coincides with Eq.(\ref{eq:stationaryT}) for the stationary state.

To understand the behavior
of $\Delta_+$ in more detail,
let us analyze the selfconsistency condition (\ref{eq:Delta_01})
for the symmetric band
of states (\ref{eq:bandwidth}). We first consider zero temperature,
when $\tanh\frac12\beta\epsilon={\rm sgn}\,\epsilon$. The integral
(\ref{eq:Delta_01}), evaluated using variable substitution
$x=4\epsilon^2$, gives
\bea
&&
\frac12\int_0^{W^2}\frac{dx}{\sqrt{(x+\Delta_+^2+\Delta_-^2)^2
-4\Delta_+^2\Delta_-^2}}
\\
&&
=\cosh^{-1}\frac{W^2+(r^2+1)\Delta_+^2}{2r\Delta_+^2}-\ln\frac1{r}
.
\eea
%
Substituting this in Eq.\,(\ref{eq:Delta_01})
and solving it, we obtain
\be\label{eq:Delta+(W)}
\Delta_+^2=\frac{W^2}{2(e^{1/\lambda}-r^2e^{-1/\lambda})\sinh\frac1{\lambda}}
,
\ee
where here and below the density of states is absorbed in the coupling,
$\lambda\nu_0\to\lambda$.
At $r=1$ we recover the BCS gap for the 
symmetric band (\ref{eq:bandwidth}), 
$\Delta_0=W/2 \sinh\frac1{\lambda}$. At $r<1$ we obtain a value 
somewhat below $\Delta_0$, the difference being small as
$(1-r^2)e^{-2/\lambda}\Delta_0$. This explains  
small departure from $\Delta_0$ seen 
in Figs.\ref{fig:gamma_temp},\ref{fig:delta_temp} at $T=0$,
as well as its absence in Fig.\ref{fig:delta_stat}.

\begin{figure}[t]
\centerline{
\begin{minipage}[t]{8in}
\vspace{-10pt}
\hspace{-40pt}
\centering
\includegraphics[width=8in]{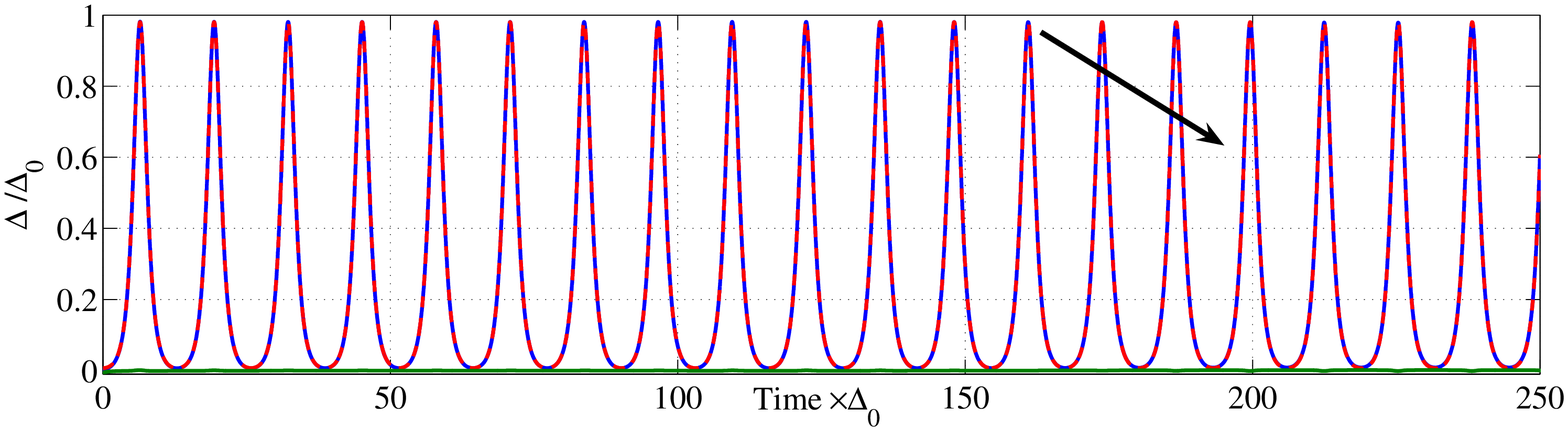}
\end{minipage}
\hspace{-2.45in}
\begin{minipage}[t]{1.6in}
\vspace{15pt}
\centering
\includegraphics[width=1.6in]{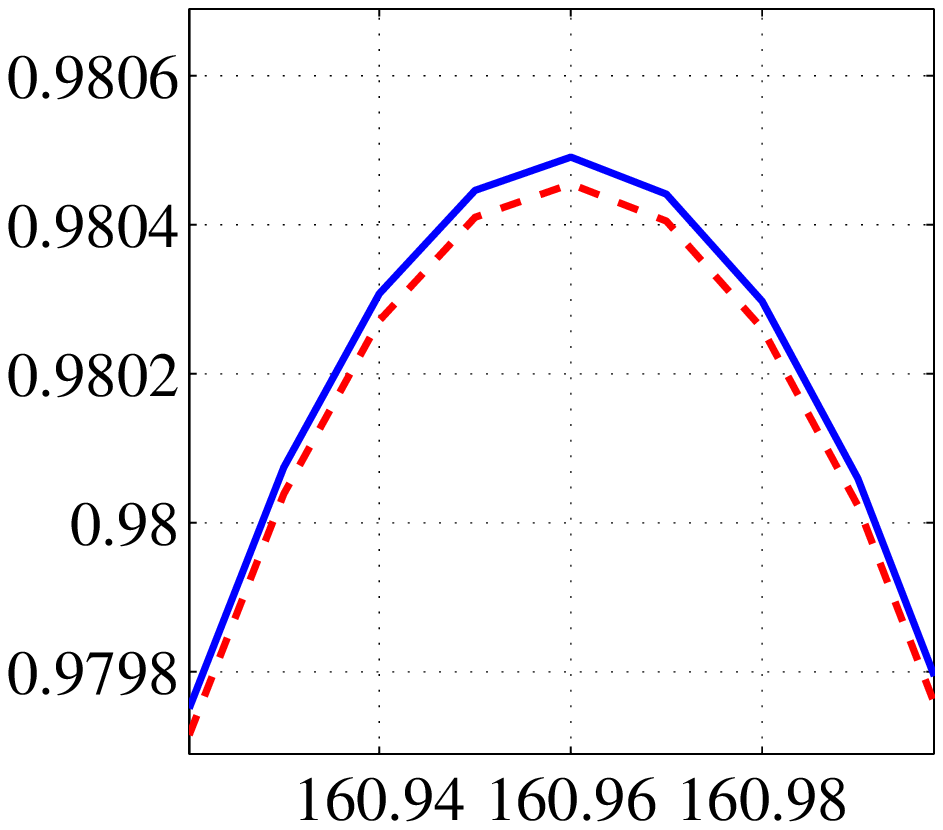}
\end{minipage}
}
\centerline{
\begin{minipage}[t]{4in}
\vspace{0.2pt}
\hspace{0.5in}
\centering
\includegraphics[width=3.3in]{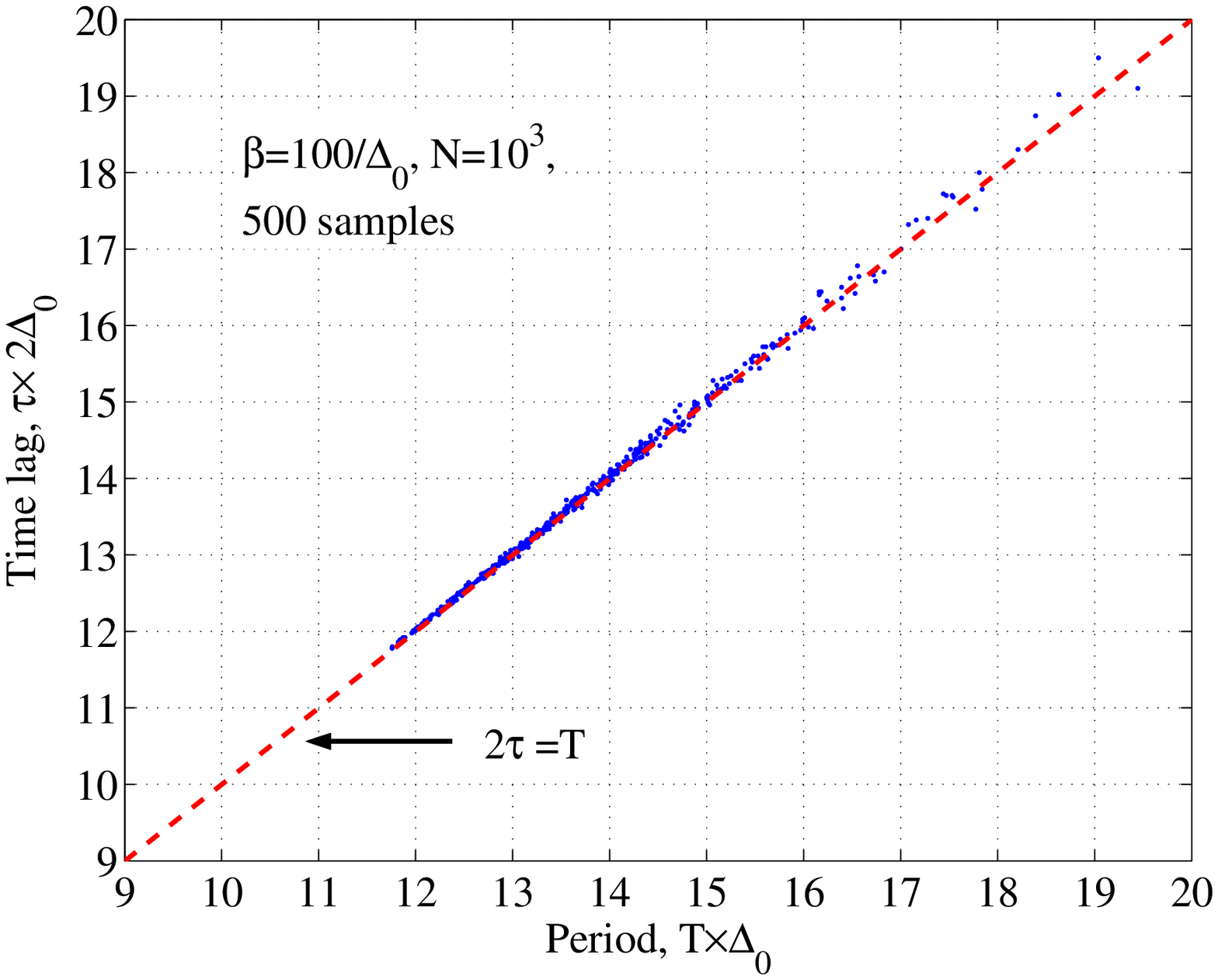}
\end{minipage}
\hspace{-0.5in}
\begin{minipage}[t]{4in}
\vspace{0.2pt}
\centering
\includegraphics[width=3.3in]{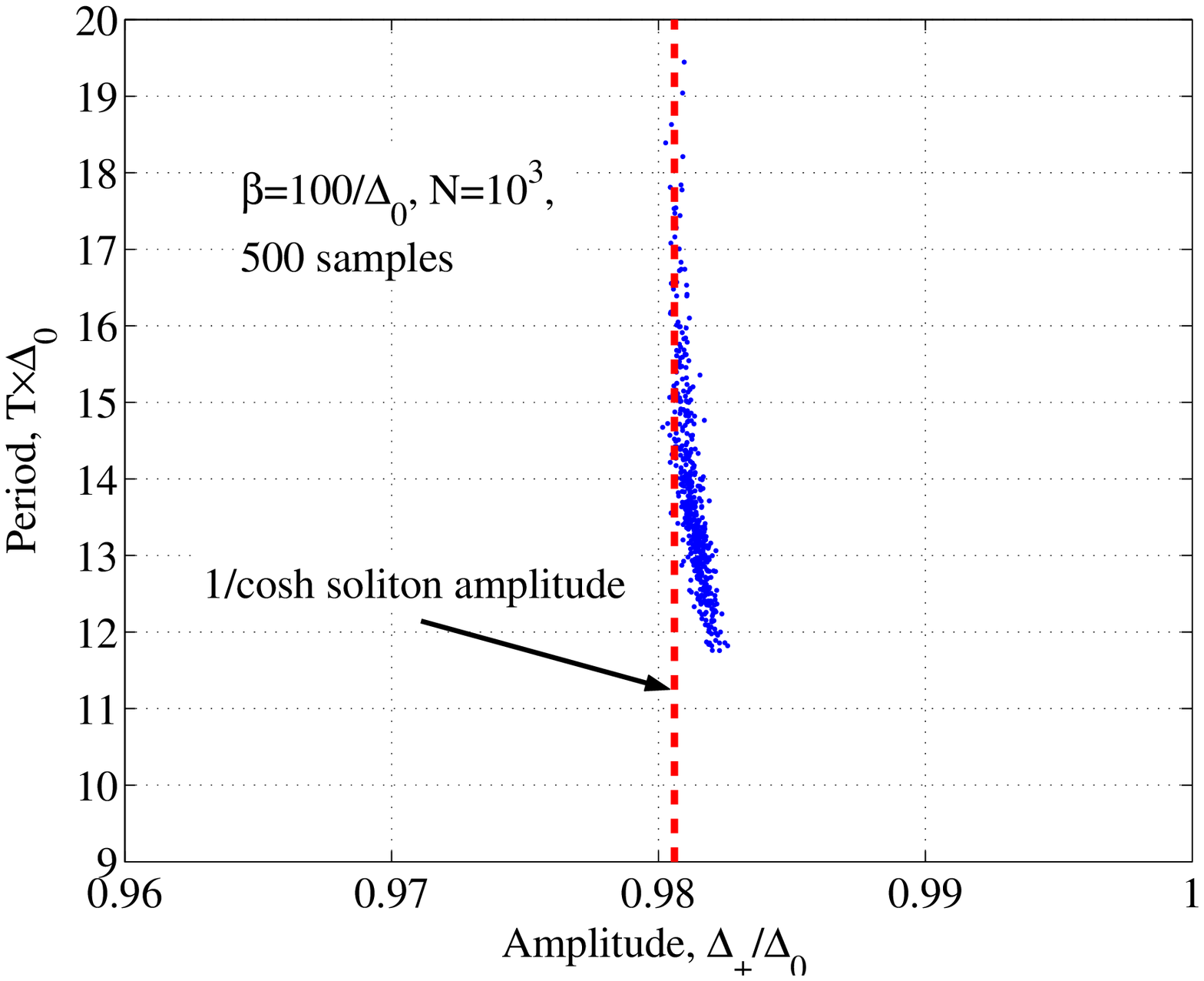}
\end{minipage}
}
\vspace{0cm}
\caption[]{
\emph{Top panel:} Comparison of the time dependence $\Delta(t)$ 
obtained from BCS/Bloch dynamics
(\ref{eq:Bloch_r}),
(\ref{eq:Gorkov})
for $N=10^3$ spins at temperature $T=10^{-2}\Delta_0$ (blue curve)
to the analytic soliton train solution (\ref{eq:Delta=dn}) 
of the same amplitude and period (green curve). 
The difference of the simulated and analytic $\Delta(t)$ 
is shown in red.
(The initial conditions (\ref{eq:r_initial}) and 
parameters $W$, $\Delta_0$ are the same as in Fig.\ref{fig:delta_sample}.)
\emph{Lower panels:} The pair distributions of the soliton train parameters for 500 different realizations:
the time lag and period (left); the period and amplitude (right).
}
\label{fig:b100}
\end{figure}

\begin{figure}[t]
\centerline{
\begin{minipage}[t]{8in}
\vspace{-10pt}
\hspace{-40pt}
\centering
\includegraphics[width=8in]{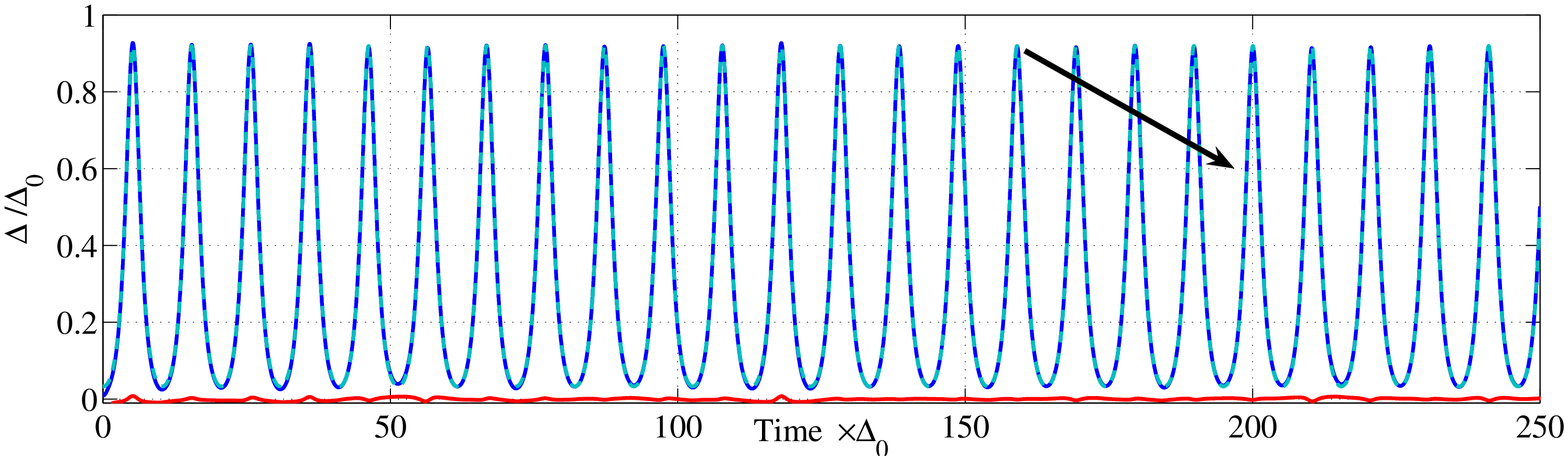}
\end{minipage}
\hspace{-2.45in}
\begin{minipage}[t]{1.6in}
\vspace{15pt}
\centering
\includegraphics[width=1.6in]{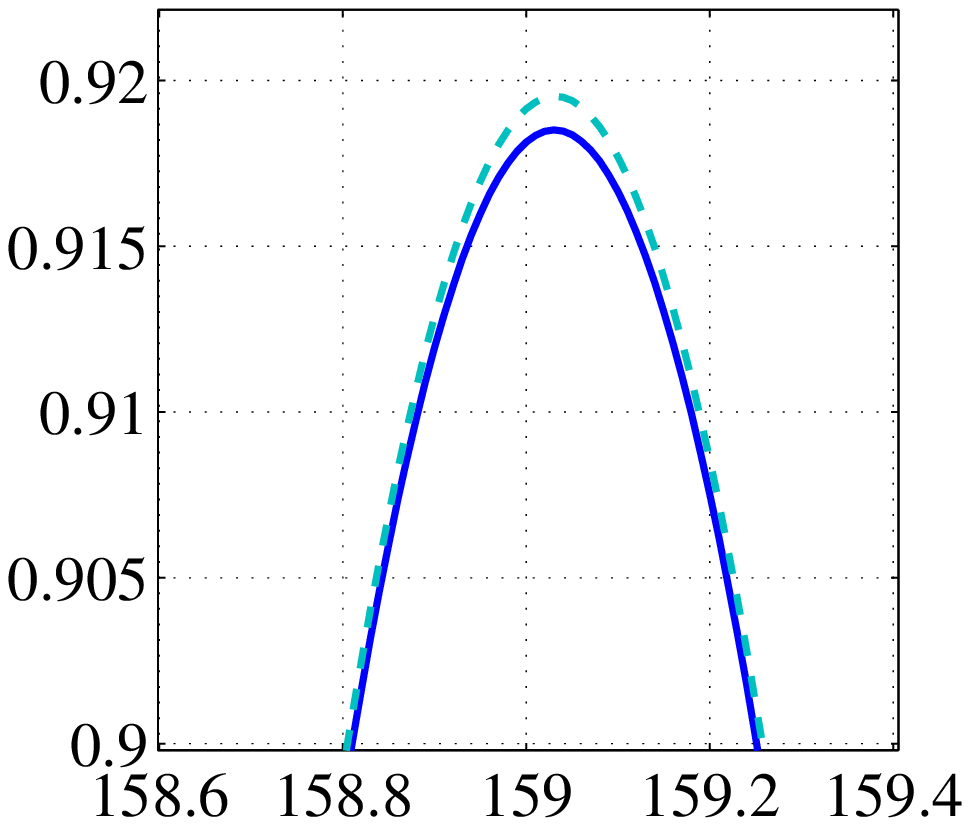}
\end{minipage}
}
\centerline{
\begin{minipage}[t]{4in}
\vspace{0.2pt}
\hspace{0.5in}
\centering
\includegraphics[width=3.3in]{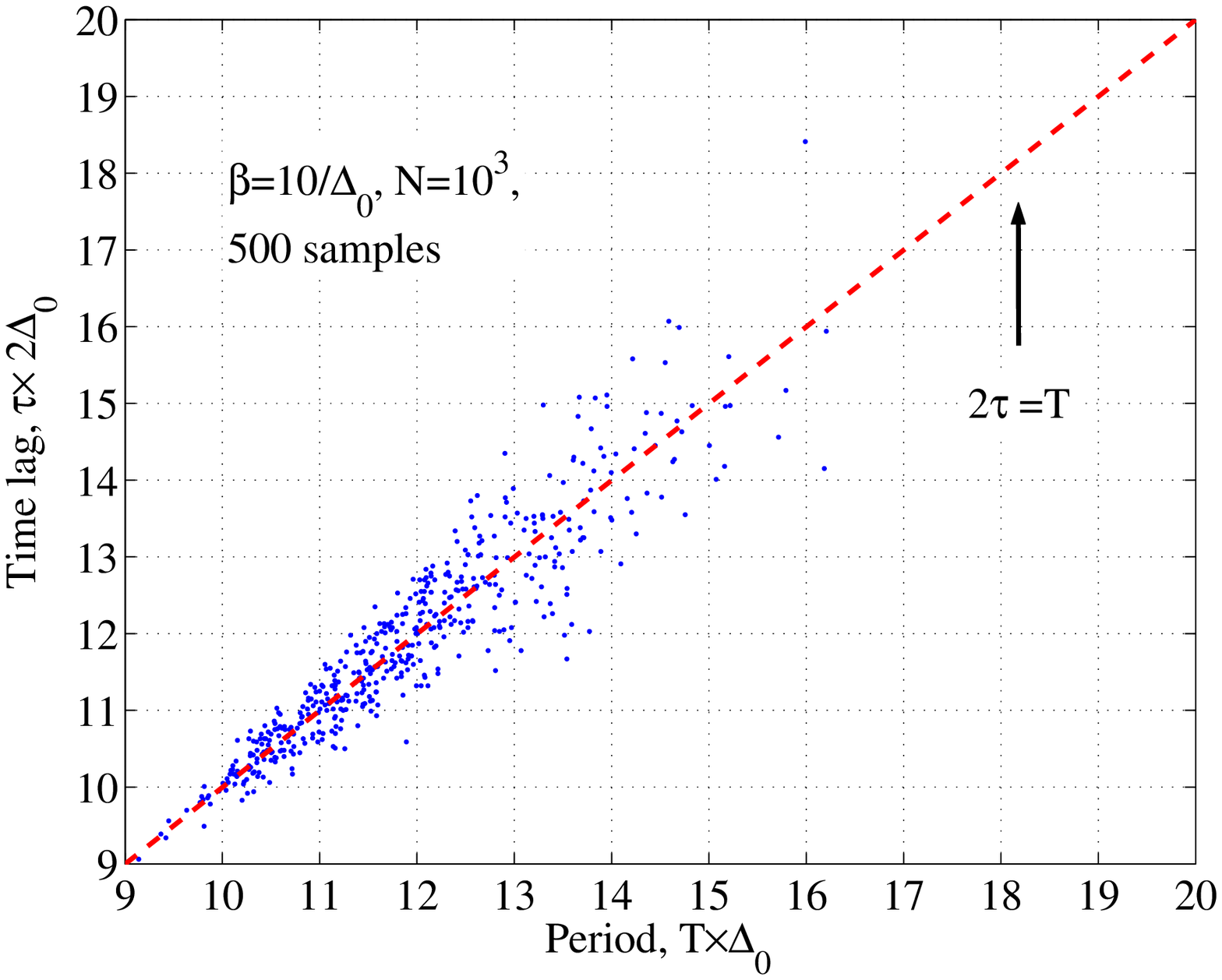}
\end{minipage}
\hspace{-0.5in}
\begin{minipage}[t]{4in}
\vspace{0.2pt}
\centering
\includegraphics[width=3.3in]{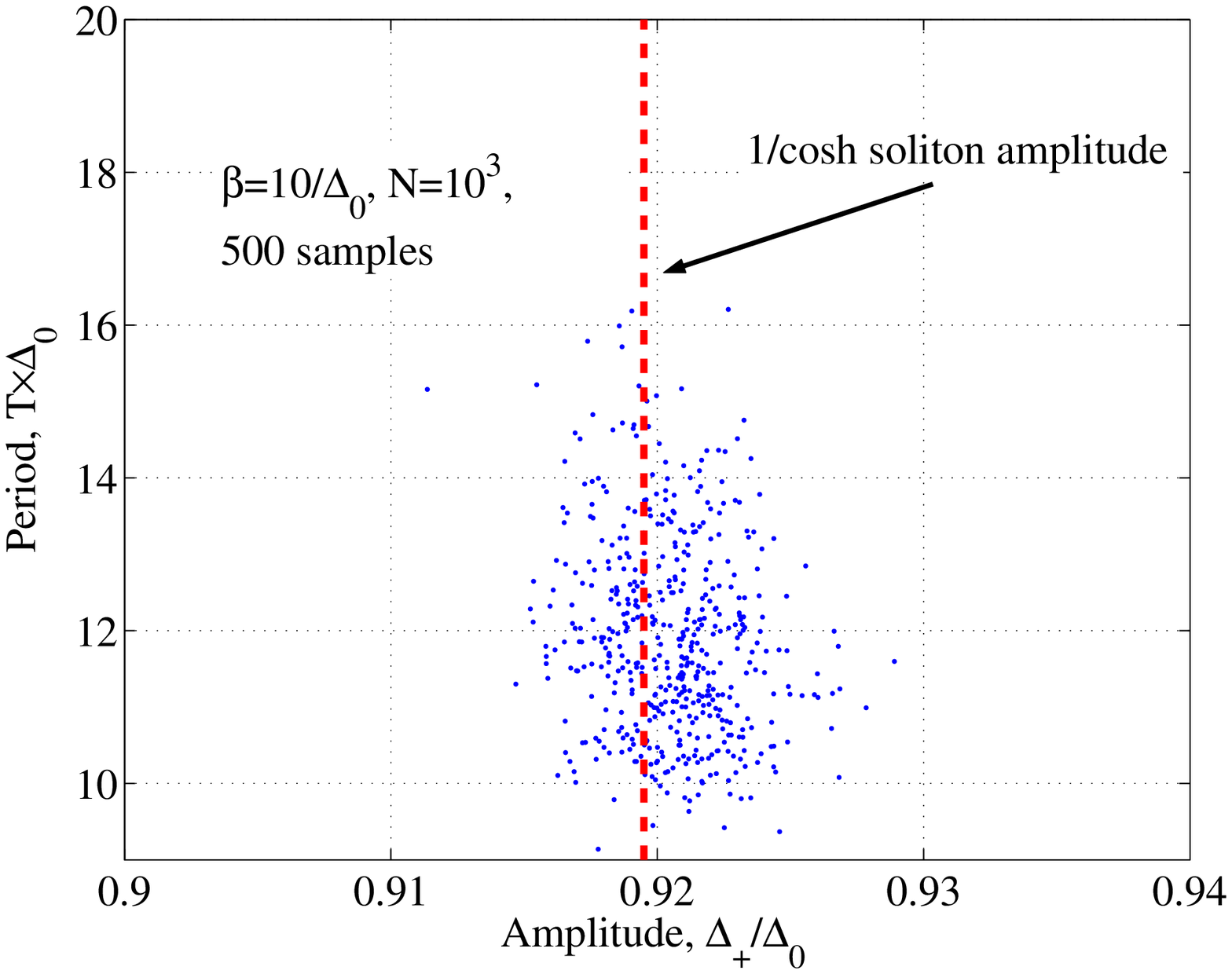}
\end{minipage}
}
\vspace{0cm}
\caption[]{
Same as in Fig.\ref{fig:b100} for higher temperature $T=10^{-1}\Delta_0$.
The simulated time dependence $\Delta(t)$ can be 
accurately fitted to the analytic solution (\ref{eq:Delta=dn}), 
with the distribution
of the period, amplitude and time lag 
somewhat broader than in Fig.\ref{fig:b100}.
}
\label{fig:b10}
\end{figure}

The simulated dynamics $\Delta(t)$ appears to be periodic (or very close to it)
over the entire time interval of the simulation.
A nearly perfect fit to $\Delta(t)$, 
as illustrated in Figs.\ref{fig:b100},\ref{fig:b10},
is provided by the elliptic function (\ref{eq:Delta=dn}) 
with the same period and amplitude.
The numerical and analytic functions are found to agree to accuracy better than
$10^{-4}$ for $\beta=100/\Delta_0$ (Fig.\ref{fig:b100} inset, top panel) 
and $10^{-3}$ for $\beta=10/\Delta_0$ (Fig.\ref{fig:b10} inset, top panel).
\breakon

\breakoff

While each realization $\Delta(t)$ is  
essentially a perfectly periodic function of time
of the form (\ref{eq:Delta=dn}), the parameters 
such as the period $T$, the time lag $\tau$, and the amplitude
$\Delta_+$ exhibit significant variations from one 
realization to another. To explore this phenomenon, we generated
a large number (500) of different realizations, and 
for each of them determined the 
values $T$, $\tau$ and $\Delta_+$ from fitting to the elliptic function
(\ref{eq:Delta=dn}). Figs.\ref{fig:b100},\ref{fig:b10} display 
the resulting pair distributions
as a set of points in the 
$(T,\tau)$ and $(T,\Delta_+)$ planes, one point per realization.

These results lead to a number of interesting observations.
First, as one would expect, 
the distribution is less noisy 
at lower temperature (Fig.\ref{fig:b100}).
Second, while at $\beta=10/\Delta_0$
the points are scattered over a 2d region, 
at $\beta=100/\Delta_0$
each distribution collapses on a 1d curve, 
indicating a specific relation between $T$, $\tau$ and $\Delta_+$
at the lower temperature.

The pair distribution of the period and the time lag 
tends to cluster around the straight line
\be
\tau = T/2
.
\ee
This can be understood from the linear stability analysis of
Sec.\ref{sec:instability}. Indeed, there we found that the linearized 
BCS problem has only two eigenvalues outside the unit circle,
$\zeta=\omega+i\gamma$ and $\zeta^\ast=\omega-i\gamma$.
The projections of the initial unpaired state on these two vectors
are close by order of magnitude. 
Now, let us take into account that 
the elliptic
function (\ref{eq:Delta=dn}) at large period $T\Delta_+\gg1$
represents a train of well-separated  
$1/\cosh$ solitons (\ref{eq:cosh_soliton}):
%
\be\label{sum_cosh}
\Delta(t)=\sum_n\frac{\Delta_+}{\cosh\Delta_+(t-t_n)}
+O(e^{-\gamma T})
,\quad
\Delta_+=\gamma
,
\ee
$t_n=nT+\tau$.
In between the solitons 
this function is given by a sum of exponential
tails of the nearest solitons:
\[
\Delta(t_n\lesssim t \lesssim t_{n+1})=
\gamma (e^{-\gamma(t-t_n)}+e^{\gamma(t-t_{n+1})})
.
\]
The amplitudes of the two terms, 
taken at some $t$ in the interval $(t_n,t_{n+1})$,
should match the $\zeta$, $\zeta^\ast$ projections
of the initial state for the numerical and analytical 
solution to coincide.
Since the terms $e^{-\gamma(t-t_n)}$, $e^{\gamma(t-t_{n+1})}$
are equal at the midpoint $t_0=\frac12(t_n+t_{n+1})$,
the time $t$ defined by the matching condition
must be close to $t_0$. This suggests that the time lag $\tau$ should indeed
be close to half a period, with the product $(2\tau-T)\gamma$ of order one
and random in sign.
This conclusion is consistent with our observations at
different temperatures (Figs.\ref{fig:b100},\ref{fig:b10}).

\breakon

\breakoff

\begin{figure}[t]
\centerline{
\begin{minipage}[t]{3.5in}
\vspace{0.2pt}
\centering
\includegraphics[width=3.4in]{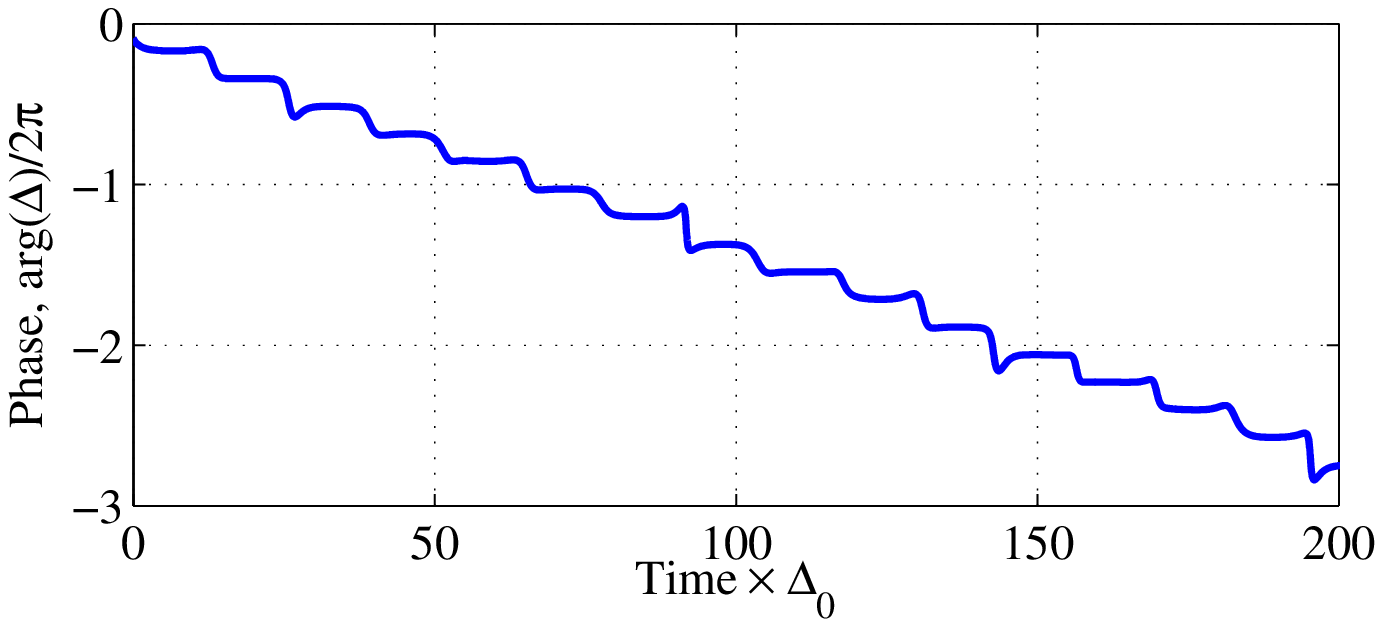}
\end{minipage}
}
\centerline{
\begin{minipage}[t]{1.9in}
\hspace{-0.2in}
\centering
\includegraphics[width=2.1in]{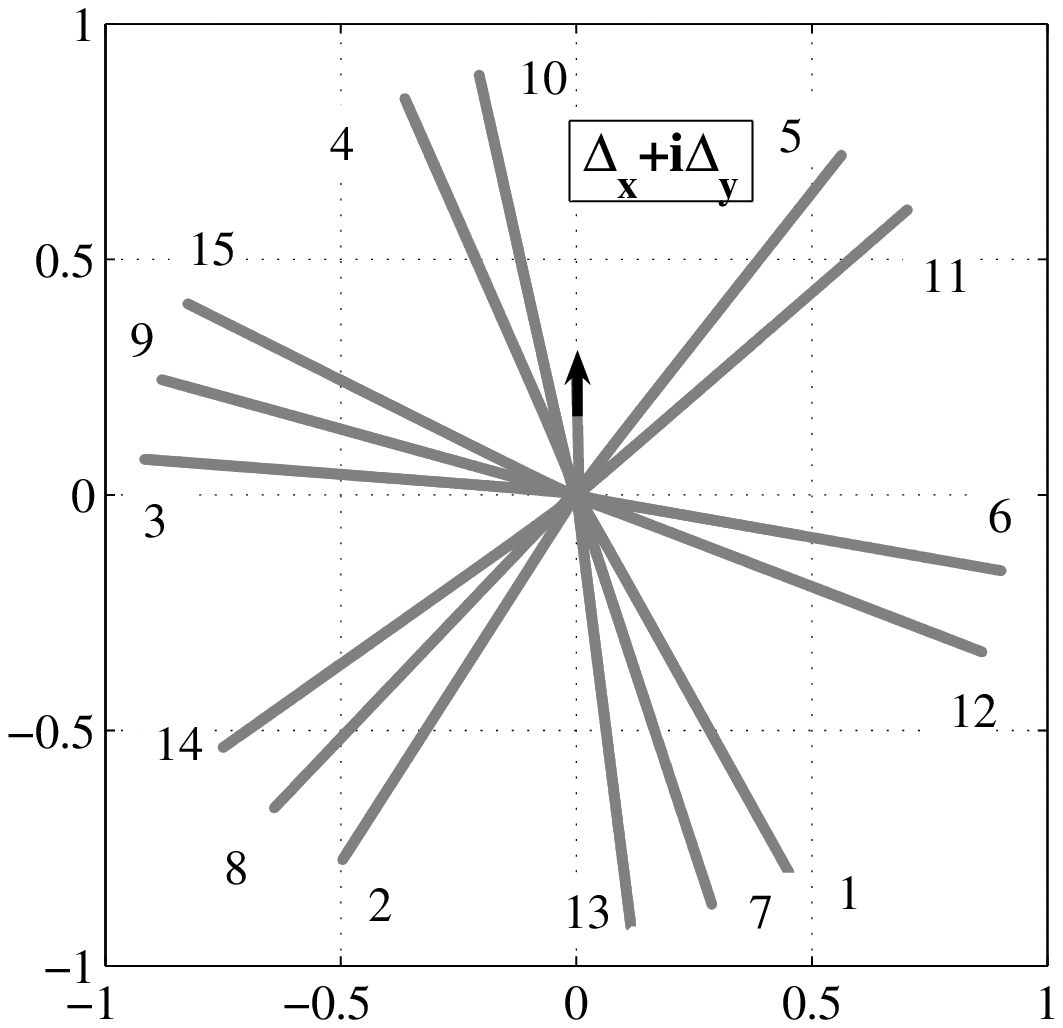}
\end{minipage}
\hspace{-0.2in}
\begin{minipage}[t]{1.9in}
\vspace{0.2pt}
\centering
\includegraphics[width=2.1in]{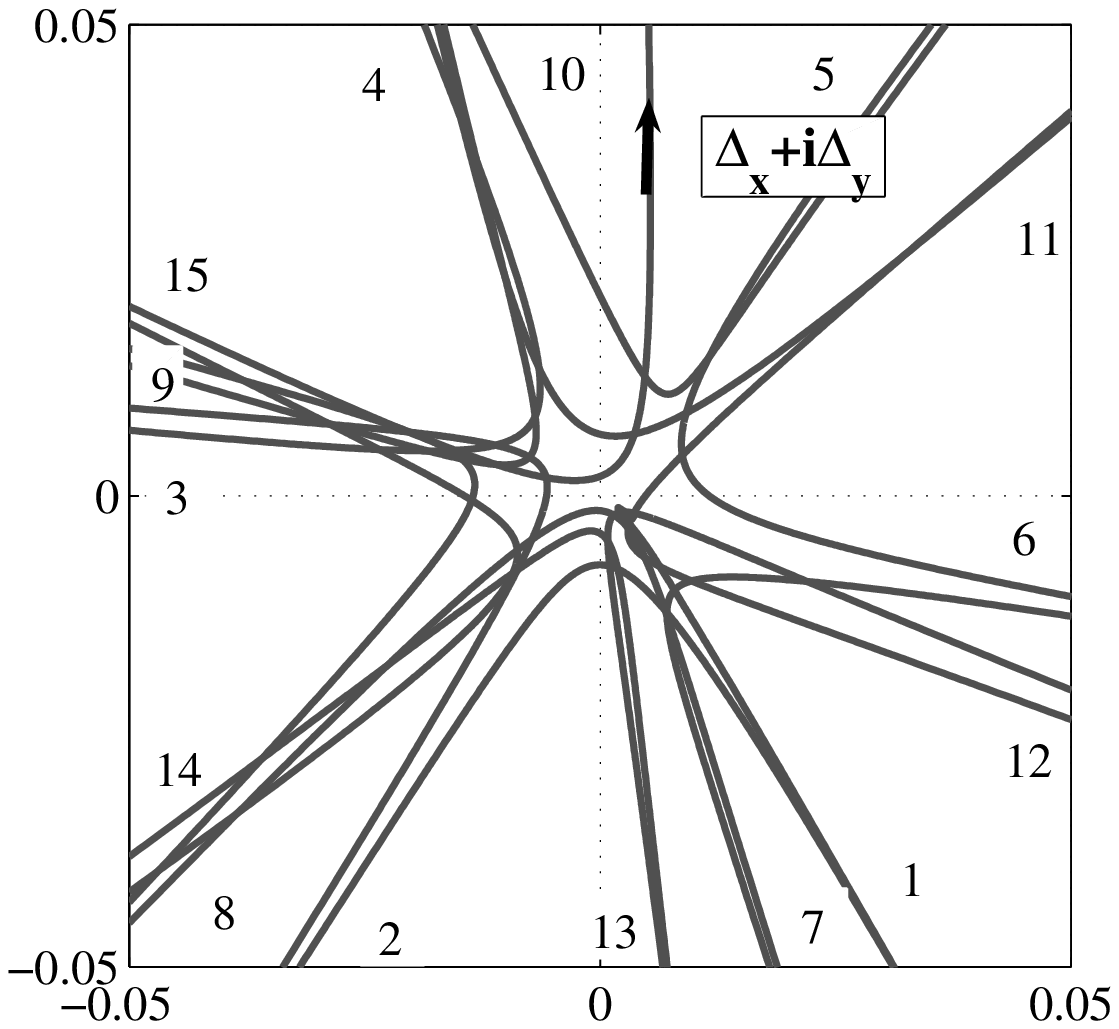}
\end{minipage}
}
\vspace{0cm}
\caption[]{
\emph{Top:} The phase of the pairing amplitude \emph{versus} time 
for the soliton train in Fig.\ref{fig:b10}.
\emph{Bottom:}
Pairing amplitude
$\Delta(t)=\Delta_x(t)+i\Delta_y(t)$ 
trajectory in the complex plane.
The phase ${\rm arg}\,\Delta(t)$ is a linear function of time
superimposed with noise.
Each radial line in the left panel corresponds to a soliton,
marked according to their order in the time sequence.
Phase shift between solitons translates into rotation by a constant angle.
The right panel shows the behavior near the origin, allowing to trace the 
order of different solitons.
}
\label{fig:phase}
\end{figure}

To complete the description of the behavior of $\Delta$, obtained 
in simulation, here we analyze the phase dynamics. The phase
$\phi={\rm arg}\,\Delta(t)$ recorded in the simulation
exhibits an approximately linear time dependence,
as illustrated in Fig.\ref{fig:phase}. As discussed above, in a system with
perfect particle-hole symmetry one expects the chemical potential to be 
pinned to the band center, in which case the condition 
$d\phi/dt=2\mu=0$ would make the phase time-independent. 
The observed linear behavior can be explained by 
particle-hole imbalance due to fluctuations of particle distribution 
$n_\vec p$ in the initial state. These fluctuations result in
nonzero chemical potential of random sign. The fluctuations 
in $\mu$
caused by random occupancy will be estimated in Sec.\ref{sec:noise},
Eq.(\ref{eq:dispersion_Tdependence}.ii). 
The magnitude and temperature dependence
of the fluctuations is found to be consistent with observations.

The noise superimposed on the linear time dependence $\phi(t)$ has several 
interesting features. First, the fluctuations about the linear dependence
show no sign of phase diffusion, since $\delta\phi$ does not grow.
Instead, they can be described as a 
periodic, or quasiperiodic, arrangement of steps connected by kinks. 
By comparing to Fig.\ref{fig:b10} 
which depicts $|\Delta|$ for this simulation, we see that each step 
is associated with a soliton, while the kinks occur in between the solitons.
This behavior can be partially understood by analyzing the trajectory
$\Delta(t)$ in a complex plane, displayed in Fig.\ref{fig:phase}.
Each radial straight line in this plot corresponds to a soliton, 
making it apparent that the phase variation occurs mainly in between the 
solitons. Interestingly, the deviation of the phase time dependence from linear
does not lead to noticeable deviation in $|\Delta|$ from the elliptic function
time dependence.

\section{Noise due to occupancy fluctuations}
\label{sec:noise}

The robustness of the elliptic function (\ref{eq:Delta=dn})
accompanied by sometimes significant variations of the parameters
$T$, $\tau$ and $\Delta_+$ among different realizations
may seem surprising.
To get insight into the origin of this behavior we consider
variations in the initial conditions, which
can be attributed to fluctuations of the pair states occupancy
(\ref{eq:Psi_bcs_T})
at $t=0$. The latter is due to thermodynamic fluctuations 
of fermion occupancy $n_\vec p$, 
and can be interpreted more intuitively as temperature fluctuations.

The first thing we note is that the existence of the soliton solutions, 
as well as their analytic form, 
is not dependent upon the details of the energy distribution $n_\vec p$,
provided the state is unpaired, 
i.e. there is no coherence in the amplitudes $u_\vec p$, $v_\vec p$ 
at different $\vec p$. 
The main difference arising for the more general distribution
is possible lack of particle-hole symmetry relative to $E_F$. In
this case, the pairing interaction shifts the chemical potential
which manifests itself as a time-dependent phase factor $e^{-i\omega t}$
multiplying $\Delta(t)$. As discussed in Ref.\cite{Barankov03}, 
this can be taken into account by a gauge transformation which shifts
single particle energies by $\omega/2$. In the transformed problem, 
only the modulus $|\Delta|$ varies with time, with its functional form
still given by the elliptic function (\ref{eq:Delta=dn}).
The parameters $\Delta_\pm$ and $\omega$ satisfy algebraic
selfconsistency equations \cite{Barankov03} with 
$\tanh \frac12\beta\epsilon_\vec p$ replaced by $1-2n_\vec p$.

The variation in the period $T$ can be linked to the fluctuations
in the initial state projection on the unstable mode
(\ref{eq:exp(gamma_t)}). Denoting this projection $\eta$, 
we can write for it a distribution of Porter-Thomas form\cite{PorterThomas},
\be
P(|\eta|=x)= 2u x \exp(-ux^2),
\ee
%
The latter
describes fluctuations of individual components of a random complex vector
in a high dimensional space, with the parameter $u$ being a function 
of the vector norm statistics. In our case, the effective dimensionality
can be estimated as a ratio of temperature to the level spacing,
$d\simeq 1/\beta\delta\epsilon=N/(\beta W)$.
To relate $\eta$ to the period $T$, we write
the time-dependent $\Delta$ at the times described by 
linear instability as $\Delta(t)\propto \eta e^{\gamma t}$.
The corresponding time range can be estimated from the condition
$\Delta(t)\lesssim\Delta_+\simeq\gamma$, giving
$t=\gamma^{-1}\ln\gamma/\eta$. 
The time $t$ is close to the 
phase lag $\tau$ which, for the reasons discussed above, is 
approximately equal to $\frac12 T$. 

Porter-Thomas distribution predicts order of magnitude
fluctuations with
typical $\eta\sim u^{-1/2}$.
This translates into fluctuations
of $T=2\gamma^{-1}\ln\gamma/\eta$ about its mean value
with the dispersion independent of $u$. 
Indeed, Figs.\ref{fig:b100},\ref{fig:b10} indicate that a 
ten-fold increase in temperature, while reducing the period, 
has little effect on its fluctuations.

To see whether the randomness in the initial state, and
specifically in the occupancy $n_\vec p$,
can explain the fluctuations in $\Delta_+$ 
recorded in Figs.\ref{fig:b100},\ref{fig:b10}, 
we consider instability of a particular initial unpaired state
(\ref{eq:Psi_bcs_T}).
Eq.(\ref{eq:instability_exponent_T}) gives 
an equation for the instability exponent:
\be\label{eq:instab_micro}
1=\lambda\sum_{\vec p}\frac{g_\vec p}{2\epsilon_{\vec p}-\zeta}
,\quad
g_\vec p=|u_{\vec p}^{(0)}|^2-|v_{\vec p}^{(0)}|^2
,
\ee
with microscopic non-averaged $u^{(0)}_{\vec p}$, $v^{(0)}_{\vec p}$
taking values zero or one with
the probabilities 
$p_{|u_{\vec p}|^2=1}=(1-n_{\vec p})^2$,
$p_{|v_{\vec p}|^2=1}= n_{\vec p}^2$ [see Eq.(\ref{eq:Psi_bcs_T})].

The fluctuation $\delta g_\vec p$ causes deviation 
$\delta\zeta$ from the average value $\zeta=i\gamma$. Linearization of 
(\ref{eq:instab_micro}) gives
\be\label{eq:delta_zeta}
\delta\zeta\sum_{\vec p}\frac{\bar g_\vec p}{(2\epsilon_\vec p-i\gamma)^2}
=-\sum_{\vec p}\frac{\delta g_\vec p}{2\epsilon_\vec p-i\gamma}
\ee
with $\bar g_\vec p=\tanh\frac12\beta\epsilon_\vec p$.
We see that both the real and the imaginary part of
$\delta\zeta=\delta\zeta'+i\delta\zeta''$ 
are nonzero, 
due to the parts of the distribution fluctuation $\delta g_\vec p$ even 
and odd relative to $E_F$. The imaginary part $\delta\zeta''$
gives fluctuation in the instability growth exponent $\gamma$. The real part
$\delta\zeta'$ can be associated with a shift of the chemical potential
due to particle-hole imbalance in the pair sector.

We estimate the magnitude of the fluctuations for low 
temperature $T\ll\Delta_0\simeq\gamma$, when 
Eq.(\ref{eq:delta_zeta}) is reduced to
\be
\delta\zeta=
\frac{i}{\gamma}\sum_{\vec p}(2\epsilon_\vec p-i\gamma)\delta g_\vec p
.
\ee
Separating the real and imaginary part, we obtain 
\be
\la\delta\zeta''^2\ra=\delta\epsilon\sum_{\vec p}\frac{4\epsilon_{\vec p}^2}{\gamma^2}
\la \delta g_{\vec p}^2\ra
,\quad
\la\delta\zeta'^2\ra=\delta\epsilon\sum_{\vec p}
\la \delta g_{\vec p}^2\ra
.
\ee
Here the second moment $\la \delta g_{\vec p}^2\ra$ is given by
\bea
\nonumber
&& \la \delta g_{\vec p}^2\ra=
\overline{(|u_{\vec p}^{(0)}|^2-|v_{\vec p}^{(0)}|^2)^2}
-(\overline{|u_{\vec p}^{(0)}|^2}-\overline{|v_{\vec p}^{(0)}|^2})^2
\\\label{eq:delta_g_ave}
&& =(1-n_\vec p)^2+n^2_\vec p-(1-2n_\vec p)^2=2n_\vec p(1-n_\vec p)
.
\eea
To obtain an order of magnitude estimate, we note that 
the fluctuations $\delta g_{\vec p}$ are of order 
one at $|\epsilon_\vec p|\lesssim T$ and exponentially
small at $|\epsilon_\vec p|\gg T$. Thus we find
\be\label{eq:dispersion_Tdependence}
{\rm (i)}\ \ \la\delta\zeta''^2\ra\simeq\frac{T^3}{\gamma^2}\,\delta\epsilon
,\quad
{\rm (ii)}\ \ \la\delta\zeta'^2\ra\simeq T\,\delta\epsilon
,
\ee
where $\delta\epsilon=W/N$ is the level spacing. 

The $T^{3/2}$ temperature dependence
of the fluctuation in $\gamma$, Eq.(\ref{eq:dispersion_Tdependence}.i), 
can be compared to the distributions
of the amplitude $\Delta_+$ presented
in Figs.\,\ref{fig:b100},\ref{fig:b10}. As we demonstrated above, 
$\Delta_+$ is numerically close to $\gamma$, becoming equal to it
in the limit of well-separated solitons
(see Fig.\ref{fig:delta_temp}). According to the $T^{3/2}$ law,
a ten-fold increase in temperature from $\Delta_0/100$ to $\Delta_0/10$
should lead to the dispersion in 
$\gamma$ increase by a factor of $30$, which is indeed  
close to the increase in 
$\Delta_+$ dispersion seen in Figs.\,\ref{fig:b100},\ref{fig:b10}.


\begin{figure}[t]
\centerline{
\begin{minipage}[t]{2in}
\vspace{0.2pt}
\centering
\includegraphics[width=1.6in]{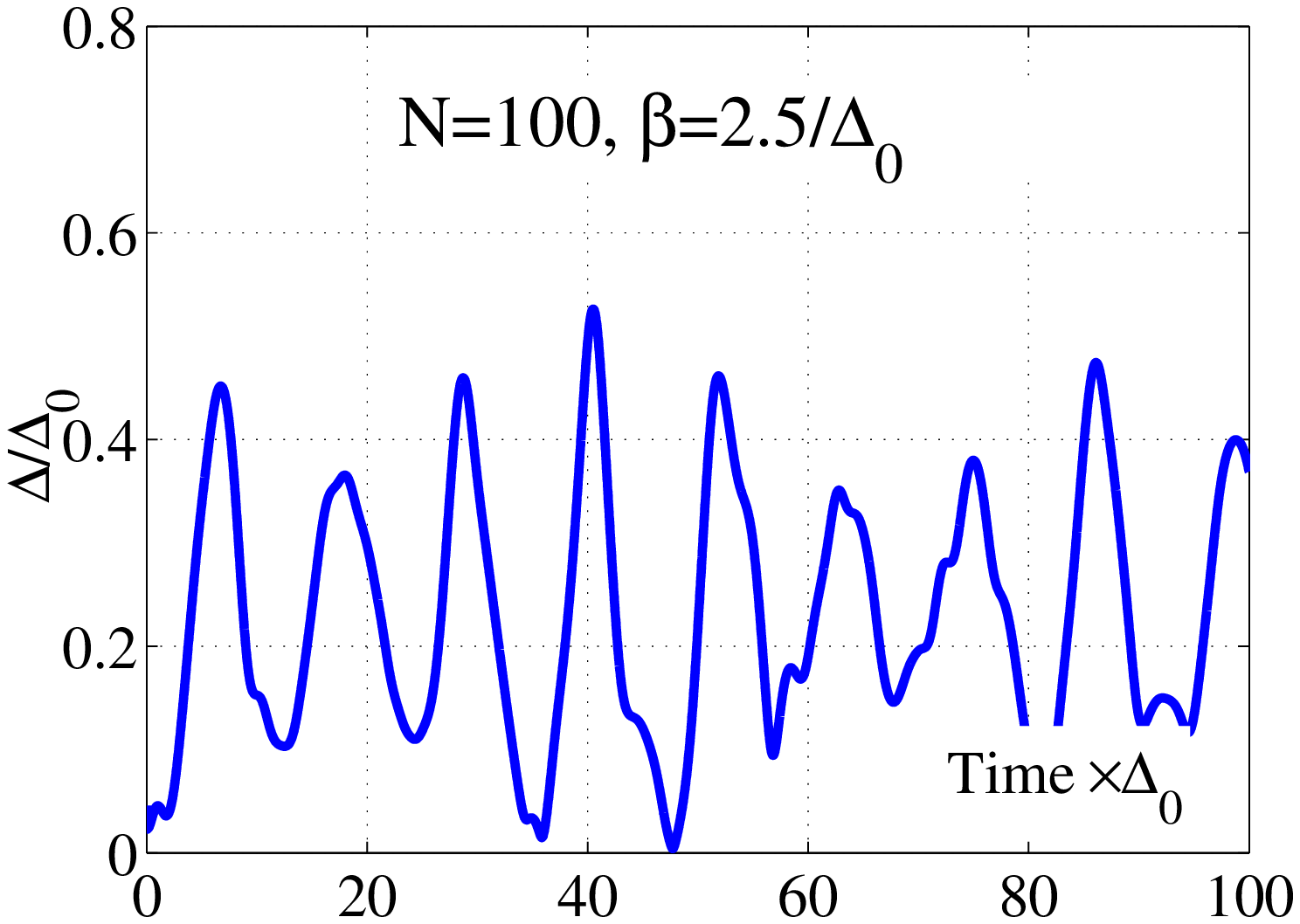}
\end{minipage}
\hspace{-0.5in}
\begin{minipage}[t]{2in}
\vspace{0.2pt}
\centering
\includegraphics[width=1.6in]{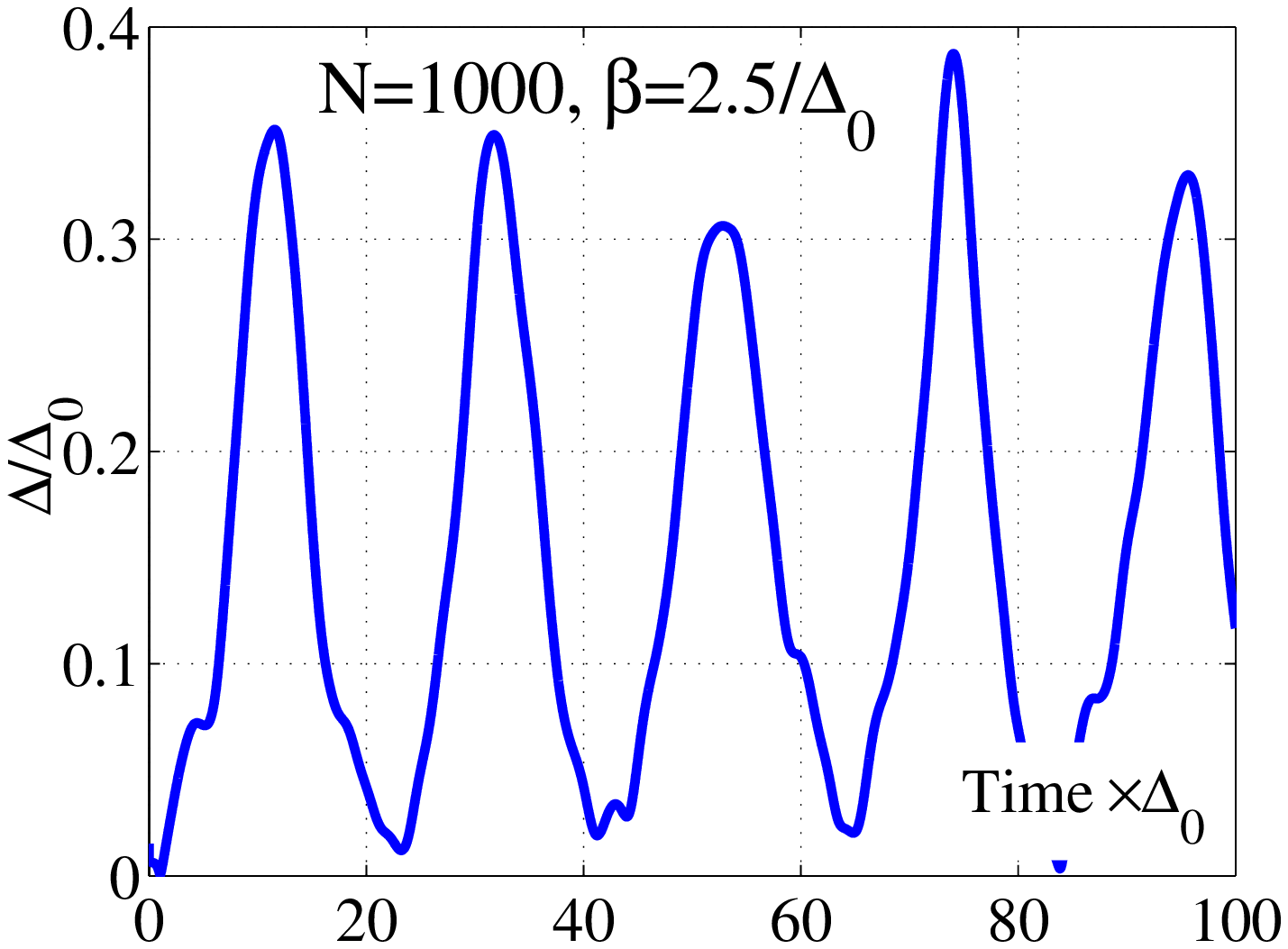}
\end{minipage}
}
\centerline{
\begin{minipage}[t]{2in}
\vspace{0.2pt}
\centering
\includegraphics[width=1.6in]{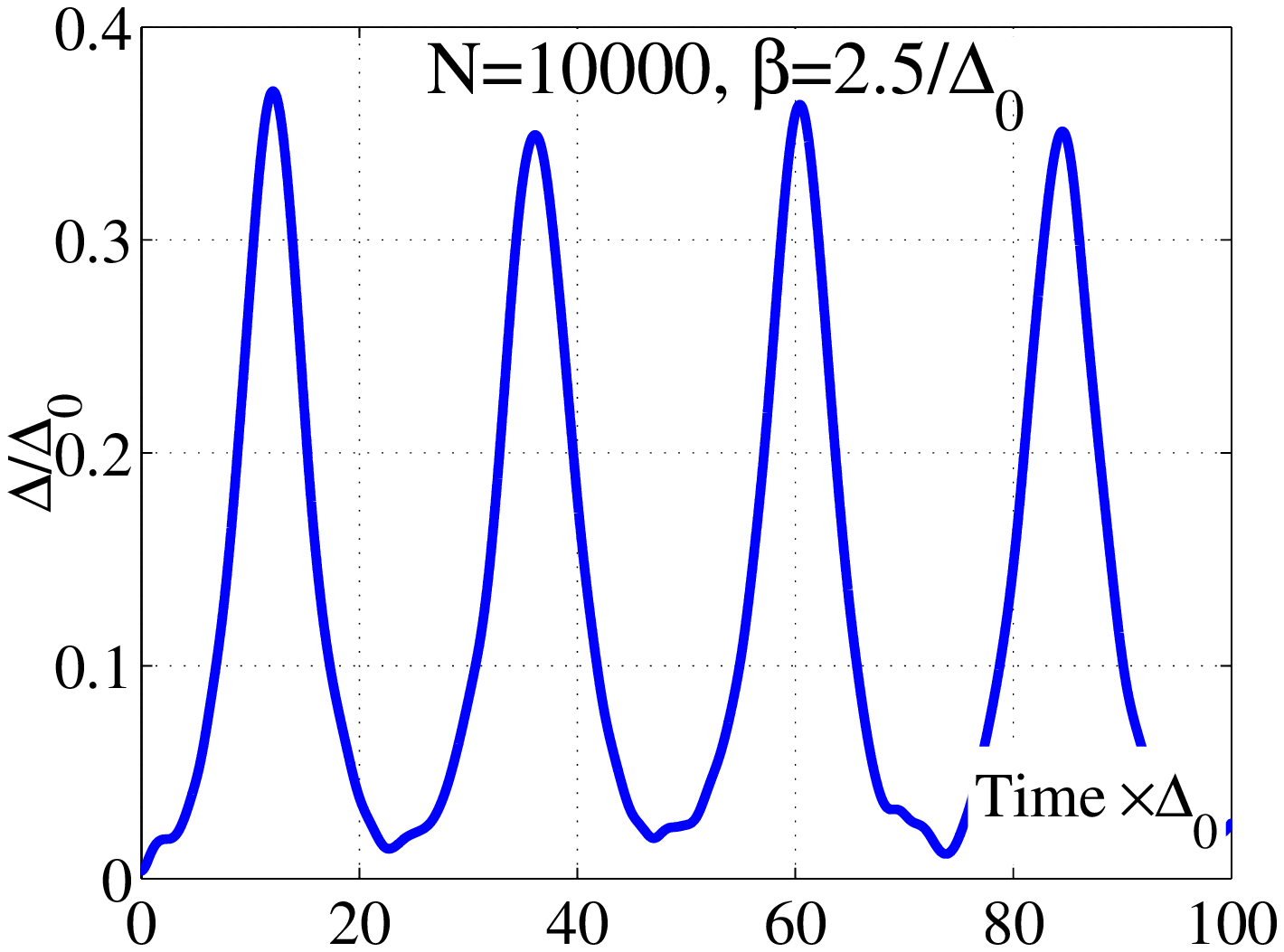}
\end{minipage}
\hspace{-0.5in}
\begin{minipage}[t]{2in}
\vspace{0.2pt}
\centering
\includegraphics[width=1.6in]{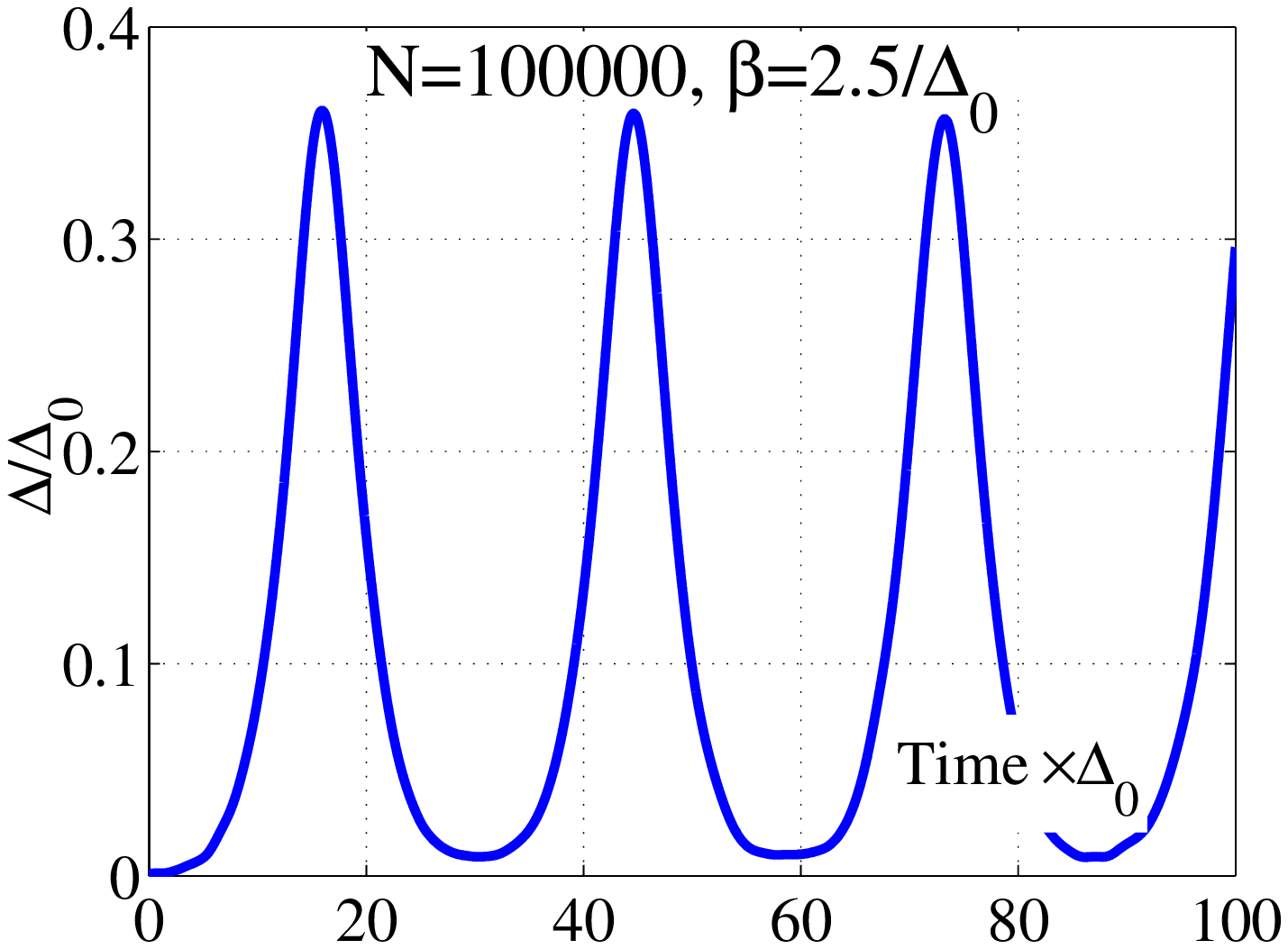}
\end{minipage}
}
\vspace{0cm}
\caption[]{
Noise suppression at increasing number of states
$N$.
The time dependence $\Delta(t)$ recorded from a simulation at
$T=0.7 T_c$ ($\beta=2.5\Delta_0$) for $N=10^2,\,10^3,\,10^4,\,10^5$
states, with other parameters the same as above. 
}
\label{fig:b2.5}
\end{figure}

\begin{figure}[t]
\centerline{
\begin{minipage}[t]{3.5in}
\centering
\includegraphics[width=3.5in]{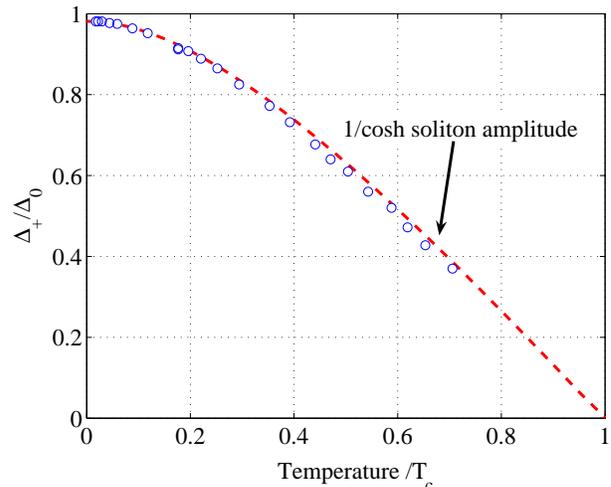}
\end{minipage}
}
\vspace{0cm}
\caption[]{
Temperature dependence of the soliton train amplitude
as recorder from the simulation. To suppress noise,
the number of levels $N$ was gradually 
increased from $N=10$ at the lowest temperature
to $N=10^5$ at the highest temperature.
Analytic fit displays
the single $1/\cosh$ soliton (\ref{eq:cosh_soliton}) amplitude
obtained from Eq.(\ref{eq:Delta_01})
at $\Delta_-=0$.
}
\label{fig:Tdependence}
\end{figure}

The noise, which quickly grows as a function of temperature, 
Eq.(\ref{eq:dispersion_Tdependence}), can be 
suppressed by reducing the level spacing $\delta\epsilon=W/N$.
The dramatic effect of level spacing on noise is illustrated in
Fig.\ref{fig:b2.5} which presents the time dependent $\Delta(t)$
at a relatively high temperature $T=0.7T_c$ for several values
of the number of levels $N$. We observe that the noise, 
at this temperature significant at small $N$, decreases
at large $N$, with the time dependence assuming the elliptic function
form (\ref{eq:Delta=dn}). As demonstrated in Fig.\ref{fig:Tdependence},
the soliton train amplitude, recorded at $N$ large enough to minimize noise,
follows very closely the analytic temperature dependence, 
Eq.(\ref{eq:Delta_01}). The value of $N$ required to reduce noise
to the level at which the behavior (\ref{eq:Delta_01}) is revealed,
grows as a function of temperature. 
While at $T=0$ as few levels as $N=10$ is quite sufficient, we find that
$N$ increases rapidly as $T_c$ is approached.

To analyze the noise at $T$ close to $T_c$, we again consider fluctuations
in the instability growth rate, given by Eq.(\ref{eq:delta_zeta}). 
At these temperatures, since $\gamma$ is linear in $T_c-T$ near $T_c$, 
we have $\gamma\ll T$. 
In this case, the sum $\sum_{\vec p}\bar g_\vec p(2\epsilon_\vec p-i\gamma)^{-2}$
in (\ref{eq:delta_zeta})
equals 
\[
4i\gamma\sum_{\vec p}
\frac{\epsilon_\vec p\bar g_\vec p}{((2\epsilon_\vec p)^2+\gamma^2)^2}
=\frac{i\pi}8 \beta
\]
After averaging over fluctuations 
$\delta g_\vec p=g_\vec p-\bar g_\vec p$, we obtain
\be
\la|\delta\zeta|^2\ra
=\lp\frac{8T}{\pi}\rp^2\delta\epsilon
\sum_{\vec p}\frac{\la\delta g_\vec p^2\ra}{(2\epsilon_\vec p)^2+\gamma^2}
.
\ee
Using the expression (\ref{eq:delta_g_ave}) for $\la\delta g_\vec p^2\ra$ 
at $|\epsilon_\vec p|\ll T$, we have
\be\label{eq:delta_zeta_Tc}
\la|\delta\zeta|^2\ra
=\lp\frac{8T}{\pi}\rp^2\delta\epsilon
\sum_{\vec p}\frac{\la\delta g_\vec p^2\ra}{(2\epsilon_\vec p)^2+\gamma^2}
=\frac{16T^2}{\pi\gamma}\delta\epsilon
.
\ee
By inspecting the right hand side of Eq.(\ref{eq:delta_zeta})
we find that the fluctuations in $\delta\zeta''$, the instability growth rate, 
dominate at $T$ close to $T_c$, 
while the fluctuations in $\delta\zeta'$, the chemical potential, 
are smaller by a factor $\gamma/T$. 
Thus Eq.(\ref{eq:delta_zeta_Tc}) gives an estimate for the fluctuations 
in $\gamma$ and, by the argument used  above, also provides an estimate of 
the noise in the soliton train amplitude $\Delta_+$.

The condition necessary for the noise (\ref{eq:delta_zeta_Tc}) 
to be small, $|\delta\zeta|\ll\gamma$, translates into
\be\label{Ncrit_Tc}
N\gg \frac{16T^2W}{\pi\gamma^3}.
\ee
We see that the minimal level number required 
to suppress noise grows as $(T_c-T)^{-3}$ near the transition. 
The fast growth is consistent with the results of simulation presented in
Figs.\,\ref{fig:b2.5},\ref{fig:Tdependence}.

The level  number $N$, which so far was taken to be arbitrary, 
can be related to other parameters as follows. For a system of size $L$ smaller
than the BCS correlation length $\xi=\hbar v_F/\Delta$,
which corresponds to a tightly trapped cold gas,
$N$ is of the order of the total particle number.
This can also be written as a relation of $N$ and particle 
Fermi momentum: $N\simeq (Lp_F/2\pi\hbar)^3$.

In an infinite system, or in a system of size larger than
$\xi$, one can define an effective $N$ equal to
the number of particles in the correlation volume,
$N\simeq (\xi p_F/2\pi\hbar)^3\approx (E_F/\Delta)^3$.
Comparing this to the inequality (\ref{Ncrit_Tc}), with the identifications
$W=E_F$, $\gamma\simeq\Delta\propto T_c-T$, we obtain a condition
$E_F^2\gg T^2$, nonrestrictive in the entire interval $0<T<T_c$. 
Other requirements for the mean field approach are also nonrestrictive:
(i) The detuning from transition,
$T_c-T$, must be outside the fluctuation region, very narrow at weak coupling;
(ii) The collisionless regime condition $\tau_\Delta\ll \tau_{\rm el}$
requires that $T_c-T$ is outside the region described by 
the time-dependent Ginzburg-Landau equation\cite{Gorkov68}, which is also
quite narrow. 
From this one can conclude that 
the mean field approach, validated by (\ref{Ncrit_Tc}), remains accurate
in an infinite system, at least for spatially uniform solutions.

It is not unconceivable that the mean field theory, known to work well
in equilibrium at weak coupling, is also valid for the dynamical problem
with generic spatially varying $\Delta$.
However, an understanding of this question can only be achieved
after the role of spatial fluctuations 
in the BCS instability is clarified\cite{Warner05}. 
Here most interesting are spatial fluctuations of the phase of $\Delta$, 
and the properties of vortices, 
which presently are not understood. If the characteristic length
scale of phase fluctuations is of order or larger than the correlation length
$\xi$, which seems likely to be the case, the dynamics of
the modulus $|\Delta|$ can be obtained in a local approximation,
using the results of this work and
ignoring spatial dependence.




\section{Spin $1/2$ representation}
\label{sec:spin_1/2}

In Sec.\ref{sec:Bloch_eqn}
we derived Bloch equations (\ref{eq:Bloch_r}) 
from Bogoliubov-deGennes equations
for $u_\vec p$ and $v_\vec p$, 
by rewriting them in the form of Gorkov equations for 
$g_\vec p=|u_\vec p|^2-|v_\vec p|^2$, $f_\vec p=2u_\vec p v^\ast_\vec p$,
and then recognizing that these quantities form a three-component Bloch vector. 
To gain more insight, here we demonstrate a different approach
in which spin $1/2$ operators and Bloch dynamics appear on an earlier stage.
Following Anderson \cite{Anderson58}, 
we define pseudospins associated with individual Cooper pair states, by 
assigning `Pauli spin' operators
$\sigma^{\pm}_{\vec p}\equiv \frac12(\sigma^x_{\vec p}\pm i\sigma^y_{\vec p})$
to each pair of
fermion states with opposite momenta as follows
\be
\label{eq:pseudospins}
\sigma^+_{\vec p}
=a^+_{\vec p\uparrow}a^+_{-\vec p\downarrow}
\,,\quad
\sigma^-_{\vec p}
=a_{-\vec p\downarrow}a_{\vec p\uparrow}
\,,
\ee
%
and $\sigma^z_{\vec p}\equiv [\sigma^+_{\vec p},\sigma^-_{\vec p}]
=a^+_{\vec p\uparrow}a_{\vec p\uparrow}
-a_{-\vec p\downarrow}a^+_{-\vec p\downarrow}$.
This allows to 
represent the BCS problem (\ref{eq:Hbcs}) as
an ensemble of interacting spins:
\be\label{eq:Hspin}
\HH={\sum}_{\vec p}'\epsilon_{\vec p}\sigma^z_{\vec p}
-2\lambda{\sum}_{\vec p,\vec q}'\sigma^+_{\vec p}\sigma^-_{\vec q}
,
\ee
where ${\sum}'_{\vec p}$ means a sum over the pairs of states
$(\vec p,-\vec p)$.
Since all the spins interact with each other equally,
the mean field theory here is exact, just like for the BCS problem.
The mean field Hamiltonian for each spin is
\be\label{eq:Hp}
\HH_{\vec p}=\vec b_{\vec p}\cdot\vec{\s_{\vec p}}
,\quad
\vec b_{\vec p}=
\lp -\Delta', - \Delta'', \epsilon_{\vec p}\rp
.
\ee
Here the $z$ component of the effective field $\vec b_{\vec p}$,
given by the single particle energy,
is spin-specific, while the transverse components,
the same for all the spins, satisfy
\be\label{eq:Delta_sigma_p}
\Delta\equiv \Delta'+i\Delta'' =
\lambda {\sum_{\vec p}}'  \la  \s^+ _{\vec p}\ra
.
\ee
This dynamical selfconsistency relation for time-dependent $\Delta$ 
and $\s^+ _{\vec p}$ is identical to the Gorkov equation (\ref{eq:Gorkov}).
In the ground state each spin is aligned with $\vec b_{\vec p}$,
and the spins form a texture near the Fermi surface
\cite{Anderson58}, given by Eq.(\ref{eq:stationary_z}),
with spin rotation described by the Bogoliubov angle.

The dynamical problem of interest takes the form
of a Bloch equation for the spins,
\be\label{eq:Bloch_sigma}
\dot \s_{\vec p} =
i[\HH_{\vec p},\s_{\vec p}]=2 \vec b_{\vec p} \times \vec \s_{\vec p}
\ee
with the field $\vec b_{\vec p}$ defined selfconsistently by
(\ref{eq:Hp}),(\ref{eq:Delta_sigma_p}).
Eq.\,(\ref{eq:Bloch_sigma}), linearized about the texture
state, describes collective excitations
of a superconductor 
with frequency $2\Delta$\cite{Volkov74,Anderson58}.
Linearized about the unpaired
state, Eq.\,(\ref{eq:Bloch_sigma}) 
describes the BCS instability  
(\ref{eq:instability_exponent}).

The Hilbert space for the spin Hamiltonian (\ref{eq:Hspin}) 
can be constructed in a standard fashion, using the states
\be\label{eq:up_down}
\{\s_\vec p\}=|...\uparrow\downarrow\downarrow\uparrow\downarrow...\ra
\ee
as basis vectors, 
where $\s_\vec p=\uparrow,\downarrow$ correspond to the fully occupied 
and empty pair states. The pair states having fermionic occupancy one 
are to be excluded as they are decoupled from the dynamics (\ref{eq:Hspin}).
The Hilbert space spanned by the states (\ref{eq:up_down}) provides 
a full representation of the Hamiltonian (\ref{eq:Hspin}).

The spin states (\ref{eq:up_down}), which are identical to the 
many-body pair states (\ref{eq:state_general}), provide the most general description of the 
problem. Here one can note, however, that the mean field
relation (\ref{eq:Delta_sigma_p}) eliminates dynamical coherence 
of different spins. This allows to simplify the state, 
replacing (\ref{eq:up_down}) by a product state
\be\label{eq:spin_product_state}
|\psi\ra=\bigotimes_{\vec p}\lp\matrix{v_\vec p\cr u_\vec p}\rp
.
\ee
Comparing this to the fermionic product states 
(\ref{eq:Psi_bcs}), (\ref{eq:Psi_bcs_T}),
we see that the spinor components $u_\vec p$, $v_\vec p$
are identical to Bogoliubov amplitudes, since the Bogoliubov-deGennes 
dynamics (\ref{eq:Bogoliubov_deGennes}) 
is equivalent to the Bloch dynamics (\ref{eq:Bloch_sigma}).

One can reduce the spin $1/2$ Bloch equations (\ref{eq:Bloch_sigma})
to the Bloch equations (\ref{eq:Bloch_r}) for classical vectors
$\vec r_\vec p$ used above as follows.
Each pair state $(\vec p,-\vec p)$ participates in the product 
(\ref{eq:spin_product_state}) with 
the probability $n^2_\vec p+(1-n_\vec p)^2$, and is excluded 
with the probability $2n_\vec p(1-n_\vec p)$. Since the Bloch equation
(\ref{eq:Bloch_sigma}) is linear in $\s_\vec p$, it takes the same form when 
$\s_\vec p$ is replaced by its expectation value
$\vec r_\vec p=\la\s_\vec p\ra$. 
One can include the probability of having occupancy
$0$ or $2$ is the expectation value, which makes the norm 
of $\vec r_\vec p$ equal $n^2_\vec p+(1-n_\vec p)^2$, in agreement with
Eq.(\ref{eq:norm_Q}). Thus we see that the spin formulation is 
indeed equivalent to the fermionic formulation employed above.



\section{Discussion}

This work demonstrates that the unpaired fermionic state, after 
being suddenly presented with pairing interaction, develops a BCS instability 
which triggers oscillations of the pairing amplitude and other quantities.
The oscillations are periodic in time and are not damped as long as 
particle collisions do not play a role. 
The oscillatory behavior comes quite naturally, 
given that without collisions the system cannot lower its energy 
to that of the BCS ground state. 

What comes as a surprise, however, is that the oscillations have 
predictable characteristics despite thermal noise in the initial conditions. 
The time-dependent pairing amplitude
is described by the soliton train solutions of  
Jacobi elliptic function ${\rm dn}$ form \cite{Barankov03}
in which only the parameters 
such as the period, amplitude and time lag depend on the initial conditions.
The explanation for such a behavior can be traced to the physics
of the BCS instability.  In the latter, when linearization over
the unpaired state is analyzed, only one mode exhibits instability, 
while other modes correspond to the perturbations that do not grow.
As a result, in the evolution of a generic unpaired fermion state
only the perturbation along the unstable direction is amplified by the 
instability, selecting the special elliptic function as a time dependence.
The accuracy to which the special solution is selected is controled
by the strength of fluctuations in the initial state, 
due to finite temperature and level spacing.

The selection phenomenon may appear counterintuitive.
Here it is instructive to make comparison 
to the results of Ref.\cite{Yuzbashyan04} which employs 
the integrability of the BCS problem to study time-dependent solutions. 
The large family of solutions obtained in Ref.\cite{Yuzbashyan04} 
could leave one under impression that they all are
equally relevant for the evolution of a generic state,
which does not agree with the results of our simulation and analytic 
arguments. Instead, as we have seen above, 
some solutions are singled out by the dynamics, while others are not. 
This peculiar situation illustrates that knowing the general solution 
of a nonlinear problem is not necessarily helpful 
in identifying the special solution 
relevant for the physical system
in which a selection mechanism is at work.

Let us also comment on some issues of interest not considered in this work.
One has to do with energy relaxation,
left out of the analysis of collisionless BCS dynamics. The relaxation
processes relevant for cold gasses are due to elastic collisions
involving two particle scattering. The rate of such processes, 
estimated above as $\tau^{-1}_{\rm el}$, 
is small compared to the typical frequency 
of oscillations $\omega\sim \Delta/\hbar$, allowing for undamped oscillations 
over a relatively long time interval $0<t\lesssim\tau_{\rm el}$. 
While proper treatment of relaxation can only be obtained with the help
of quantum kinetic theory, 
one can account for it heuristically\cite{Barankov03}
by inserting a Landau-Lifshitz term in the Bloch equation (\ref{eq:Bloch_r}),
changing $\vec b_\vec p$ to 
\[
\vec b'_\vec p=\vec b_\vec p-\frac1{\tau_{\rm el}}\vec l_\vec p\times\vec r_\vec p
,
\]
where $\vec l_\vec p=\vec b_\vec p/|\vec b_\vec p|$. The resulting 
evolution exhibits damped oscillations converging to the BCS ground state
asymptotically at large times $t\gg \tau_{\rm el}$.

Another phenomenon of interest is related with spatial fluctuations.
Our discussion focused on the dynamics in a system of finite size,
of order or smaller than the correlation length,
in which we considered spatially uniform $\Delta(t)$, neglecting
the spatial dependence altogether. In an infinite system,
one expects the emergent pairing dynamics 
to exhibit phase fluctuations and vortices simultaneously with the 
oscillations of the modulus $|\Delta|(t)$ similar to that studied above.
If the phase fluctuations occur at distances larger 
than the correlation length and the vortices are dilute, 
one can use the estimates made at the end 
of Sec.\ref{sec:noise} to argue that the modulus $|\Delta|$ will oscillate
in a pretty much the same way as in the spatially uniform
situation. However, more work will be needed to clarify this. 

\section{Summary}

In this paper we studied fermionic pairing in a system with time-dependent 
pairing interaction. 
We analyze the situation when the pairing builds up after
the interaction has been abruptly turned on. Theoretical analysis,
supported by numerical simulations, predicts a stage of exponential growth, 
described by BCS instability of the unpaired Fermi gas, 
followed by periodic oscillations described by collisionless 
nonlinear BCS dynamics. We consider spatially uniform 
situation relevant for systems of small size, and find that:

(i) 
In the collisionless approximation, 
at times shorter than the energy relaxation time,
the oscillations are undamped;

(ii) 
The time dependence of the pairing amplitude is obtained from
an exact solution of the nonlinear BCS problem
\cite{Barankov03}, 
a periodic soliton train described by 
the Jacobi elliptic function ${\rm dn}$, with parameters 
depending on the microscopic initial conditions;
 
(iii)
The robustness of the elliptic function behavior is explained by 
a dynamical selection process, in which the BCS instability acts 
to amplify the initial perturbation in a specific unstable 
mode of the system, 
generating a time dependence with predictable characteristics;

(iv) 
The fluctuations of the amplitude, period and time lag 
of the soliton train can be accounted for 
by occupation probability fluctuations in the initial state.

\acknowledgements

This work benefited from discussions with B. Spivak.

\appendix
\section{} 

Here we present yet another derivation of the Bloch equation 
(\ref{eq:Bloch_r}), starting from the evolution of time-dependent
amplitudes $u_{\vec p}(t)$, $v_{\vec p}(t)$. Let us consider the 
Bogoliubov-deGennes equation (\ref{eq:Bogoliubov_deGennes}) 
with the selfconsistency condition (\ref{eq:Delta_BCS}).
By introducing a new   
variable $w_{\vec p}=u_{\vec p}/v_{\vec p}$, 
the pair of linear differential equations
(\ref{eq:Bogoliubov_deGennes}) is reduced to a single nonlinear equation
of Riccati form,
\be\label{eq:w_p}
i\p_t w_{\vec p}
= 2\epsilon_{\vec p} w_{\vec p} + \Delta(t) - \Delta^\ast(t) w_{\vec p}^2
\ee
which was analyzed in Ref.\cite{Amin04}.
The selfconsistency condition (\ref{eq:Delta_BCS}), rewritten
in terms of $w_{\vec p}(t)$, becomes
\be\label{eq:Delta_w}
\Delta(t)=
\lambda\sum_{\vec p}Q_{\vec p}\frac{w_{\vec p}(t)}{1+|w_{\vec p}(t)|^2}
,
\ee
where 
\be
Q_{\vec p}=|u_\vec p|^2+|v_\vec p|^2
\ee
is the norm of $(u_\vec p, v_\vec p)$, 
conserved by the dynamics (\ref{eq:Bogoliubov_deGennes}). 

The initial value $w_{\vec p}$ corresponding to the unpaired Fermi gas
(\ref{eq:u0v0_T})
is $w_{\vec p}^{(0)}=e^{i\phi_\vec p}(1-n_\vec p)/n_\vec p
=e^{\beta\epsilon_\vec p+i\phi_\vec p}$,
with the phases $\phi_\vec p$ random and uncorrelated for different $\vec p$.
For $(u_\vec p, v_\vec p)$ of the form (\ref{eq:u0v0_T}) we have
\[
Q_{\vec p}=n^2_\vec p+(1-n_\vec p)^2
\]
The factor $Q_{\vec p}<1$ describes the effect of Pauli blocking 
which was discussed in Sec.\ref{sec:Bloch_eqn}.

\begin{figure}[t]
\centerline{
\begin{minipage}[t]{3.5in}
\centering
\includegraphics[width=2.0in]{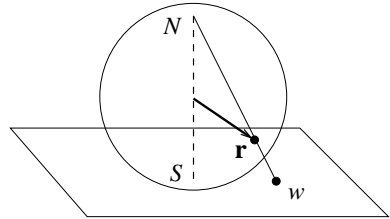}
\end{minipage}
}
\vspace{0cm}
\caption[]{
Stereographic projection (\ref{eq:stereographic}) schematic.
}
\label{fig:stereographic}
\end{figure}

The next step is to perform a reverse stereographic projection
of the complex variable $w$, mapping it onto the unit sphere
$r_1^2+r_2^2+r_3^2=1$ as follows:
\be\label{eq:stereographic}
r_1+ir_2=\frac{2w}{|w|^2+1}
,\quad
r_3=\frac{|w|^2-1}{|w|^2+1}
\ee
(see Fig.\ref{fig:stereographic}).
The dynamics (\ref{eq:w_p}), written in terms of $z=r_1+ir_2$ and $r_3$,
gives
\be\label{eq:z,r3}
\frac{dz}{dt}=-2i\epsilon_\vec p z+2i\Delta r_3
,\quad
\frac{dr_3}{dt}=i\Delta^\ast z-i \Delta z^\ast
.
\ee
After rewriting Eq.(\ref{eq:z,r3})
in terms of $\vec r_{\vec p}=(r_1,r_2,r_3)_\vec p$,
we again obtain the Bloch
equation (\ref{eq:Bloch_r}), 
$\dot\vec r_{\vec p}=2\vec b_\vec p\times \vec r_{\vec p}$,
with the ``magnetic field''
$\vec b_\vec p=-(\Delta',\Delta'',\epsilon_\vec p)$.

The selfconsistency condition (\ref{eq:Delta_w}) takes the form
\be\label{eq:selfconsistencyT}
\Delta=\frac{\lambda}{2}\sum_{\vec p}Q_{\vec p}z_{\vec p}
.
\ee
The difference in the form of the selfconsistency relations 
(\ref{eq:selfconsistencyT} and (\ref{eq:Gorkov}) is due to
the difference in normalization of $\vec r_\vec p$ 
here and in Sec.\ref{sec:Bloch_eqn}. Here we have
$|\vec r_\vec p|=1$, while in Sec.\ref{sec:Bloch_eqn} 
we had $|\vec r_\vec p|=n^2_\vec p+(1-n_\vec p)^2<1$
which is precisely the factor $Q_{\vec p}$ needed to account for the difference 
in the norm.



\section{} 

Here we estimate the direct heating of the Fermi system due to
the shakeup caused by interaction switching
\[
\HH_{\rm int}(t)=\lambda(t)\sum_{p_1+p_2=p_3+p_4}a^+_{\vec p_4,\alpha}a^+_{\vec p_3,\beta}a_{\vec p_2,\beta}a_{\vec p_1,\alpha}.
\]
The interaction time dependence during switching is step-like, 
varying from $0$ to $\lambda$ over a characteristic time $\tau_0$, so 
the Fourier component $\lambda_\omega=\int e^{i\omega t}\lambda(t)dt$
bah Aves as $\omega^{-1}$ at $\omega\tau_0\ll 1$.
For example, the interaction switching model
$\lambda(t)=\lambda(1-e^{-t/\tau_0})\theta(t)$
gives
$\lambda_\omega=i\lambda/(\omega(1-i\omega\tau_0))$.

The effective temperature $T_{\rm eff}$ after switching can be
estimated from the Golden Rule transition rate for 
particle energy excitation matched by
the net energy increase and fermion specific heat:
\bea\label{eq:Teff}
&& a T_{\rm eff}^2=
\sum_{\omega,\,1...4}
\hbar\omega
n_1n_2(1-n_3)(1-n_4)|\lambda_\omega|^2
\times
\\\nonumber
&&
\qquad\qquad
\delta(\hbar\omega-\epsilon_{\vec p_3}-\epsilon_{\vec p_4}+\epsilon_{\vec p_1}+\epsilon_{\vec p_2})
\eea
with $n_i=n_{\vec p_i}$ the occupation numbers of the states $\vec p_{1...4}$, 
$a=\frac{\pi^2}6\nu$.
At $T=0$, we obtain
$a T_{\rm eff}^2=\sum_{\omega>0} \hbar\omega N_\omega |\lambda_\omega|^2$
with $N_\omega=\frac1{6}\nu^4\omega^3$.


The integral over $\omega$ gives $T_{\rm eff}^2 \simeq \lambda^2\nu^3\tau_0^{-3}$.
Comparing $T_{\rm eff}$ to $\Delta_0$ we find that the system is not overheated,
$T_{\rm eff}\ll\Delta_0$,
when
$E_F\tau_0\gg (\lambda n/\Delta_0)^{2/3}$.
This condition is compatible with the
nonadiabaticity requirement $\tau_0\ll \tau_{\Delta}$,
allowing for a range of possible values of the switching times $\tau_0$
for which heating of the Fermi gas is negligible.

\end{multicols}
\end{document}